\documentclass[11pt,a4paper]{article}
\usepackage[T1]{fontenc}
\usepackage[utf8]{inputenc}
\usepackage{amsmath,amssymb,mathtools}
\usepackage{graphicx,booktabs}
\usepackage{mathptmx}
\usepackage[margin=1in]{geometry}
\usepackage{setspace}
\usepackage{authblk}
\usepackage[colorlinks=true, linkcolor=magenta, urlcolor=blue, citecolor=blue]{hyperref}
\allowdisplaybreaks[2]
\numberwithin{equation}{section}
\newcommand{\dd}{\mathrm{d}}
\newcommand{\ee}{\mathrm{e}}
\newcommand{\ii}{\mathrm{i}}
\newcommand{\rc}{r_{\rm crit}}
\newcommand{\Imag}{\operatorname{Im}}

\title{\textbf{A unified dark sector from a single canonical scalar field in Bohmian Wheeler--DeWitt cosmology with a cuscuton-inspired potential}}
\author[1]{Anant Badal\thanks{anantbadal22@iitk.ac.in}}
\author[1]{Rikpratik Sengupta\thanks{rikpratiks@iitk.ac.in}}
\author[1]{Kaushik Bhattacharya\thanks{kaushikb@iitk.ac.in}}

\affil[1]{Department of Physics, Indian Institute of Technology Kanpur, Kalyanpur 208016, India}
\date{}

\begin{document}
\maketitle

\begin{abstract}
\noindent We quantize a spatially flat Friedmann--Lema\^itre--Robertson--Walker minisuperspace containing one canonical scalar field with the hyperbolic potential generated by a cuscuton-type superpotential, and extract cosmological histories from the Wheeler--DeWitt wave function in the de Broglie--Bohm interpretation. For the canonical scalar field potential parameter, related to the cuscuton couplings in the original cuscuton model, below an explicit critical value the potential has a unique nondegenerate minimum with positive height, and the ratio \(\kappa\) of the scalar mass to the de Sitter Hubble rate at the minimum diverges as \(\kappa^2\simeq 3b/\delta\), where \(\delta\) is the relative distance to criticality. For the original hyperbolic potential, \(\kappa\leq 3/\sqrt{2}\), whereas its cuscuton-inspired generalization allows a large hierarchy between the scalar oscillation frequency and the expansion rate. For \(\kappa\gg1\), small excitation and sufficiently large cell action, the reduced wave equation has an adiabatic Gaussian solution describing a de Sitter vacuum, a classical oscillating condensate whose averaged excitation density redshifts as \(a^{-3}\), and, at the packet centre, a leading quantum density equal to the zero-point energy of the effective homogeneous oscillator divided by the physical cell volume, which is also dust-like. Full numerical solutions of the Wheeler--DeWitt equation support the averaged zero-point law at the reported \(10^{-3}\) level, reproduce its dependence on the position of the Bohmian particle inside the packet, show that quantum corrections to the present Hubble rate converge as the inverse of the cell action, and support a single-field dust-plus-vacuum description at the homogeneous background level when \(\kappa\gg1\). The classical histories approach the \(\Lambda\)CDM distance relations with a deviation of \(1.3/\kappa\) in \(D_V\). The dark energy component has \(w=-1\) by construction, and the model is not fitted to data.
\end{abstract}

\section{Introduction}

The late-time universe is dominated by two components whose microscopic nature remains unknown. A cosmological constant, or something that mimics it closely, accounts for the accelerated expansion, and a pressureless component of unknown origin accounts for the clustering of matter and for the growth of structure. Within the standard cosmological model the two are introduced independently, with energy densities that coincide in order of magnitude only at the present epoch. The smallness of the vacuum energy relative to any natural microscopic scale is the cosmological constant problem \cite{Weinberg1989,Martin2012}, and the nature of the dark matter is constrained by a large body of cosmological and astrophysical data but is not identified \cite{Planck2018}. Recent baryon acoustic oscillation measurements from the Dark Energy Spectroscopic Instrument \cite{DESI2024,DESI} have sharpened the observational situation, since combinations of these data with supernovae and the cosmic microwave background are reported to prefer a dark energy component whose density varies with time. Whatever the final interpretation, the expansion history at redshifts below two is now measured at the percent level, and any model that describes the dark sector must be confronted with distances at that precision.

A long tradition seeks a unified description in which a single field supplies both the vacuum energy and the pressureless matter. Fluid models of the Chaplygin type \cite{Kamenshchik2001,Bilic2002,Bento2002} and purely kinetic scalar models \cite{Scherrer2004} realize this idea through a specific equation of state, and a review of the scalar field versions can be found in Ref.~\cite{Bertacca2010}. A second class of models uses a canonical scalar field with a potential that has a positive minimum \cite{Sahni2000,Matos2000,Arbey2006}. Around the minimum the field oscillates, and when the oscillation frequency greatly exceeds the Hubble rate the period-averaged energy density decays as \(a^{-3}\) with vanishing average pressure \cite{Turner1983}. The height of the minimum then acts as a cosmological constant, and the oscillating component is a coherent condensate of scalar quanta, which is the picture underlying ultralight and fuzzy dark matter \cite{Hu2000,Hui2017,Marsh2016,Ferreira2021}. The structure of this construction contains a requirement that is rarely stated explicitly. The ratio of the curvature scale of the potential at its minimum to the Hubble rate set by the height of the minimum must be large. In most models this hierarchy is a property of the chosen parameters and has no structural explanation.

Quantum cosmology provides an independent route to the same problem, since the quantum state of the universe may modify the effective background even when the classical potential is fixed. In the minisuperspace truncation the Wheeler--DeWitt equation \cite{DeWitt1967} becomes a wave equation for a few homogeneous variables, whose solutions have been studied extensively with different boundary conditions \cite{Hartle1983,Vilenkin1984,Kiefer1988,Halliwell1991,Kiefer2012}. The interpretation of the wave function is the delicate point, because the equation has no external time and the conserved current is not positive definite. The de Broglie--Bohm interpretation \cite{Bohm1952,Holland1993,Durr1992} resolves this by supplementing the wave function with an actual configuration whose motion is guided by the phase, which defines a single expansion history without reference to an observer. The approach has been developed in detail for cosmology \cite{Colistete2000,Peter2007,Peter2008,PintoNeto2013,Kiefer2022} and has been used to remove singularities \cite{Falciano2007,Demaerel2019}. A recent Bohmian minisuperspace model by Basilakos, Kouniatalis, Saridakis and Tzerefos \cite{Bqc1} uses a canonical transformation to bring the Hamiltonian of a scalar field with a hyperbolic potential into that of a two-dimensional hyperbolic oscillator, which makes the Wheeler--DeWitt equation separable. We show in Sec.~\ref{sec:potential} that their solvable parameter choice has \(\mu_\star/H_\star=\sqrt{9/8}\), below the threshold \(3/2\) above which a damped scalar oscillates, so its classical scalar perturbations near the minimum are overdamped and do not supply a rapidly oscillating matter component in that regime.

The cuscuton \cite{Afshordi2006,Afshordi2007} suggests another source of potentials. In the one-field first-order construction of Bazeia, Dantas and Santos da Costa \cite{CuscDE} the scalar sector is generated by a superpotential \(W(\phi)=A\cosh(B\phi)-\alpha_{\rm c}\phi\) in which the cuscuton coupling \(\alpha_{\rm c}\) deforms a hyperbolic potential and lowers its minimum. We use this potential in a canonical minisuperspace model. The kinetic term is canonical and the square-root cuscuton action is not quantized, so the question we ask is what a deformation of this type does to the dark-sector structure of the quantum model. The answer turns out to be sharper than expected. The deformation is the only ingredient that produces the scale separation required for a dust-like condensate, and it does so through a simple critical behaviour of the vacuum.

The main results can be summarized as follows. The domain of couplings with a positive vacuum energy is determined in closed form, with a critical coupling at which the vacuum energy vanishes, and near that coupling the mass-to-Hubble ratio at the minimum diverges as \(\kappa^2\simeq3b/\delta\). For \(\kappa\gg1\) the Wheeler--DeWitt equation reduces to a time-dependent oscillator whose Gaussian solutions are known exactly, and the centred Bohmian trajectories in these solutions have a leading quantum density that scales as dust and equals the zero-point energy of the effective homogeneous oscillator divided by the physical cell volume. Direct integration of the full wave equation reproduces this law, its dependence on the position of the guided configuration within the packet, and the approach of the tested expansion histories to their classical counterparts as the cell action grows. In the classical limit the histories approach those of \(\Lambda\)CDM with a deviation in \(D_V\) that decreases as \(1.3/\kappa\). We work throughout at the level of the homogeneous background with a single scalar degree of freedom and a fixed operator ordering, so perturbations, baryons and radiation are not part of the discussion, and the parameters of the quantum state are specified as inputs and are not derived. Within this framework the results are exact where stated and numerically controlled elsewhere.

The paper is organized as follows. Section~\ref{sec:minisuperspace} sets up the quantum minisuperspace model and the Bohmian observables. Section~\ref{sec:potential} analyzes the potential and its vacuum. Section~\ref{sec:adiabatic} derives the adiabatic Gaussian solution and the quantum dust law. Section~\ref{sec:numerics} presents the full numerical solution and its tests. Section~\ref{sec:observables} translates the results into late-time distance observables. In section~\ref{sec:semiclassical_limit} an approximate method of producing the classical limit from the Bohmian framework has been presented and Sec.~\ref{sec:conclusions} summarizes and discusses the outlook. The proofs and numerical methods are collected in the appendices.

%%%%%%%%%%%%%%%%%%%%%%%%%%%%%%%%%%%%%%%%%%%%%%%%%%%%%%%%%%%%%%%%%%%%%%%%%%%%%%%%%%%%%%%%%%%%%%%%%%
\section{Quantum minisuperspace and Bohmian guidance}
\label{sec:minisuperspace}

We use the signature \((-,+,+,+)\) and units \(c=\hbar=8\pi G=1\) unless stated otherwise. The spatially flat line element with a lapse function is \(\dd s^2=-N^2(t)\dd t^2+a^2(t)\,\dd\mathbf x^2\), and the matter content is a homogeneous canonical scalar field \(\phi(t)\) with potential \(U(\phi)\). With an arbitrary lapse, the Hubble parameter is \(H=\dot a/(Na)\), where the dot denotes a derivative with respect to the coordinate time \(t\). In the cosmic-time gauge \(N=1\), used for the evolution equations below, this reduces to \(H=\dot a/a\). Integrating the Einstein--Hilbert action and the scalar action over a comoving cell of coordinate volume \(V_0\), and discarding a total derivative, one obtains the reduced Lagrangian
\begin{equation}
    {\mathcal L}=V_0\left[-\frac{3a\dot a^2}{N}+\frac{a^3\dot\phi^2}{2N}-Na^3U(\phi)\right].
    \label{eq:lagrangian}
\end{equation}
The canonical momenta are \(p_a=-6V_0a\dot a/N\) and \(p_\phi=V_0a^3\dot\phi/N\), and the Legendre transform gives \({\mathcal H}=N\mathcal C\) with
\begin{equation}
    \mathcal C=-\frac{p_a^2}{12V_0a}+\frac{p_\phi^2}{2V_0a^3}+V_0a^3U(\phi).
\end{equation}
The lapse is a Lagrange multiplier, and its variation imposes \(\mathcal C\approx0\). The kinetic term of the scale factor has the opposite sign to that of the scalar field, so the supermetric on the two-dimensional minisuperspace has Lorentzian signature, which is the origin of the hyperbolic character of the Wheeler--DeWitt equation. In terms of \(\beta=\ln a\), with \(p_\beta=ap_a\), and after multiplication of the constraint by the positive factor \(V_0a^3\), the constraint becomes
\begin{equation}
    -\tfrac1{12}p_\beta^2+\tfrac12p_\phi^2+V_0^2\,\ee^{6\beta}\,U(\phi)\approx0 .
    \label{eq:constraint}
\end{equation}
Classically the rescaling is accompanied by the lapse redefinition \(\tilde N=N\ee^{-3\beta}/V_0\) and leaves the constraint surface and the trajectories unchanged. In the quantum theory the choice of which form of the constraint to quantize is part of the ordering ambiguity \cite{Hawking1986}, and a family of orderings is obtained by adding a term proportional to \(\partial_\beta\Psi\). We quantize the form in Eq.~\eqref{eq:constraint} with no first-derivative term, and the dependence on the ordering is not explored here.

It is convenient to measure the potential in units of its minimum value. We write \(U=\Lambda_\star F(x)\) with \(x=B\phi\) for a field-scale constant \(B\), where \(F\) is normalized so that its minimum equals one and \(\Lambda_\star=3H_\star^2\) is the corresponding vacuum energy density, and we introduce the logarithmic scale factor \(u=\ln(a/a_{\rm i})\) relative to an initial slice \(a_{\rm i}\). The Wheeler--DeWitt equation depends on one dimensionless combination of the cell volume and the vacuum Hubble rate, which we call the cell action,
\begin{equation}
    \nu=6V_{\rm i}H_\star,\qquad V_{\rm i}=V_0a_{\rm i}^3,\qquad c_x^2=6B^2\equiv6b .
    \label{eq:nu}
\end{equation}
Promoting \(p_u\to-\ii\partial_u\) and \(p_x\to-\ii\partial_x\) in Eq.~\eqref{eq:constraint} gives
\begin{equation}
    \Psi_{uu}-c_x^2\Psi_{xx}+\nu^2\ee^{6u}F(x)\Psi=0 .
    \label{eq:wdw}
\end{equation}
This is a Klein--Gordon equation on the two-dimensional minisuperspace, with \(u\) playing the role of a time-like coordinate, \(c_x\) the characteristic speed and \(\nu^2\ee^{6u}F\) a position-dependent squared mass that grows with the scale factor. Restoring the units, \(\nu\ee^{3u}=6VH_\star/(8\pi G\hbar)\) for a physical cell volume \(V\), so the cell action is the only place where \(\hbar\) enters, and the classical limit is \(\nu\to\infty\) at fixed classical data. For a cell of the size of the Hubble volume \(\nu\) is of order \(10^{121}\), which indicates how far the physical regime lies from the values of \(\nu\) used in numerical work below. With Dirichlet conditions on \(x\) the quantity \(J=\int\Imag(\Psi^*\Psi_u)\,\dd x\) is conserved, and it provides a global diagnostic of the numerical solutions. It is not a probability, since the integrand has no definite sign, and the Bohmian description does not require it to be one.

In the de Broglie--Bohm interpretation the wave function is written in polar form \(\Psi=R\,\ee^{\ii S}\), and the canonical momenta are identified with the gradients of the phase. Combining this with the classical relations between velocities and momenta in the gauge \(N=1\) one obtains the guidance equations. We denote the Hubble rate in units of its vacuum value by \(h=H/H_\star\) and the field velocity by \(v=\dd x/\dd(H_\star t)\), and the guidance equations read
\begin{equation}
    h=-\frac{\ee^{-3u}}{\nu}\,S_u,\qquad v=\frac{c_x^2\ee^{-3u}}{\nu}\,S_x ,
    \label{eq:guidance}
\end{equation}
where the expanding branch is selected by \(S_u<0\). These equations determine the trajectory \(x_{\rm B}(u)\) once an initial configuration is given, and the proper time follows from \(\dd\tau/\dd u=1/h\), with \(\tau=H_\star t\). The configuration \(x_{\rm B}\) at the initial slice is an input that is not fixed by \(\Psi\). Because Eq.~\eqref{eq:wdw} is second order in \(u\) and its conserved current is not positive, the usual quantum-equilibrium argument that distributes the configuration with density \(|\Psi|^2\) is not available in a straightforward form, and we do not assume any distribution. The guided configuration is instead placed at the centre of the initial packet or displaced by a fixed multiple of its width, and the dependence of the results on this choice is tested explicitly in Sec.~\ref{sec:numerics}. Using \(u\) as the evolution variable requires \(h\ne0\), which holds along all expanding histories discussed here.

Substituting the polar form into Eq.~\eqref{eq:wdw} and separating real and imaginary parts gives a Hamilton--Jacobi equation with a quantum correction and a continuity equation,
\begin{align}
    S_u^2-c_x^2S_x^2&=\nu^2\ee^{6u}F+\frac{R_{uu}-c_x^2R_{xx}}{R},\label{eq:HJ}\\
    \partial_u(R^2S_u)-c_x^2\partial_x(R^2S_x)&=0 .\label{eq:continuity}
\end{align}
Dividing Eq.~\eqref{eq:HJ} by \(\nu^2\ee^{6u}\) and using Eq.~\eqref{eq:guidance} gives the effective Friedmann constraint
\begin{equation}
    h^2=\frac{v^2}{c_x^2}+F+Q,\qquad Q=\frac{\ee^{-6u}}{\nu^2}\,\frac{R_{uu}-c_x^2R_{xx}}{R},
    \label{eq:friedmann}
\end{equation}
in which \(K=v^2/c_x^2\) and \(F\) are the kinetic and potential energies of the scalar field in units of \(\Lambda_\star\), and \(Q\) is the quantum density in the same units. Because the quantum term involves \(R_{uu}\) as well as \(R_{xx}\) it differs from the quantum potential of nonrelativistic mechanics, and a packet that moves quickly in minisuperspace has a quantum density that is reduced by a Lorentz-like factor. Equation~\eqref{eq:friedmann} is an identity once \(Q\) is computed from the phase, and substituting \(Q\) obtained that way back into the constraint is not an independent test of the wave equation. We therefore compute \(Q\) from the amplitude when we wish to check the solution, and from the phase when we need it as a diagnostic.

The Bohmian equation of motion for the field follows by differentiating Eq.~\eqref{eq:HJ} with respect to \(x\) and using Eq.~\eqref{eq:guidance}, and it contains the quantum force \(-3bQ_x/h\) in addition to the classical force (Appendix~\ref{app:bohm}),
\begin{equation}
    \frac{\dd v}{\dd u}=-3v-\frac{3b\,(F_x+Q_x)}{h}.
    \label{eq:bohm-eom}
\end{equation}
Differentiating the constraint~\eqref{eq:friedmann} along the trajectory and using Eq.~\eqref{eq:bohm-eom}, together with \(\dd F/\dd u=F_xv/h\) and \(\dd Q/\dd u=(\partial_uQ)_x+Q_xv/h\), the terms containing \(Q_x\) cancel and one finds
\begin{equation}
    \frac{\dd h}{\dd u}=-\frac{3K}{h}+\frac{(\partial_uQ)_x}{2h}.
    \label{eq:raych}
\end{equation}
The derivative in the last term is taken at fixed \(x\). If it were replaced by the total derivative along the trajectory, an erroneous term proportional to \(Q_x\,\dd x_{\rm B}/\dd u\) would appear. The geometric equation of state and the deceleration parameter are
\begin{equation}
    w_{\rm tot}=-1-\frac{2}{3h}\frac{\dd h}{\dd u},\qquad q_{\rm dec}=\frac{1+3w_{\rm tot}}2,
    \label{eq:wtot}
\end{equation}
and with Eq.~\eqref{eq:raych} the total equation of state takes the form \(w_{\rm tot}=(K-F+\Pi_Q)/h^2\), where
\begin{equation}
    \Pi_Q=-Q-\tfrac13(\partial_uQ)_x
    \label{eq:PiQ}
\end{equation}
is the effective quantum pressure in units of \(\Lambda_\star\). The scalar and quantum components exchange energy, although their sum obeys the total continuity equation, and for this reason \(w_{\rm tot}\) and the canonical scalar equation of state \(w_\phi=(K-F)/(K+F)\) differ for a Bohmian history. In the classical limit \(Q\to0\) the system reduces to
\begin{equation}
    \frac{\dd x}{\dd u}=\frac vh,\qquad\frac{\dd v}{\dd u}=-3v-\frac{3bF_x}{h},\qquad\frac{\dd h}{\dd u}=-\frac{v^2}{2bh},
    \label{eq:classical}
\end{equation}
with \(h^2=v^2/c_x^2+F\), which are the Friedmann, Klein--Gordon and Raychaudhuri equations written in \(u\) and \(H_\star\) units. Equation~\eqref{eq:classical} for \(\dd h/\dd u\) is valid only classically, and for a quantum trajectory \(\dd h/\dd u\) must be obtained from Eq.~\eqref{eq:raych} or directly from the phase, as described in Sec.~\ref{sec:numerics}.

%%%%%%%%%%%%%%%%%%%%%%%%%%%%%%%%%%%%%%%%%%%%%%%%%%%%%%%%%%%%%%%%%%%%%%%%%%%%%%%%%%%%%%%%%%%%%%%%%%
\section{The cuscuton-inspired potential and its vacuum}
\label{sec:potential}

The cuscuton is a scalar field with a square-root kinetic term, with action
\[
    S=\int d^4x\,\sqrt{-g}\left[\alpha_{\rm c}\sqrt{|\partial_\mu\phi\,\partial^\mu\phi|}-V(\phi)\right]
\] 
in the notation of Ref.~\cite{Afshordi2006}, whose cosmological background equations are first order and which carries no local propagating degree of freedom when it is the only scalar. In the present case \(\alpha_c\ge 0\) is the non-negative deformation parameter.

The scalar potential employed in this work is inspired by the cuscuton-based superpotential construction of Ref.~\cite{CuscDE}. In that construction, a superpotential of the form
\[
    W(\phi)=A\cosh(B\phi)-\alpha_c\phi
\]
leads to the potential
\begin{eqnarray}
    V(\phi)=
    \frac32[A\cosh(B\phi)-\alpha_c\phi]^2
    -\frac12A^2B^2\sinh^2(B\phi).
    \label{eq:V1}
\end{eqnarray}
In the present work, we adopt this functional form as the potential of a purely canonical scalar field, \(U(\phi)=\lambda_1V(\phi)\), and, at the classical level, the scalar field satisfies
\begin{eqnarray}
    \ddot{\phi} + 3H\dot{\phi} + U_\phi(\phi)=0\,,
\end{eqnarray}
in our paper, where \(U_\phi\) implies the derivative of the potential with respect to the field. The cuscuton kinetic interaction and the first-order dynamical relations of the original construction are not imposed. Thus, the parameter \(\alpha_c\) serves only as a deformation parameter controlling the shape of the canonical scalar potential.

We take \(U=\lambda_1V\) with a positive effective coefficient \(\lambda_1\). The reference \cite{CuscDE} uses \(4\pi G=1\) and Ref.~\cite{Bqc1} uses \(8\pi G=1\), and a change of the gravitational normalization rescales the Einstein--Hilbert term and the whole matter action, so it cannot in general be undone by multiplying the potential alone. Only the shape of the potential is imported, and \(\lambda_1\) is a free scale that is absorbed in the vacuum Hubble rate. In the variables \(x=B\phi\), \(b=B^2\) and \(r=\alpha_{\rm c}/(AB)\) one has \(U=\lambda_1A^2f(x)\) with
\begin{equation}
    f(x)=\tfrac32(\cosh x-rx)^2-\tfrac b2\sinh^2x,\qquad f_x=(3-b)\sinh x\cosh x-3r(\cosh x+x\sinh x)+3r^2x .
    \label{eq:f}
\end{equation}
The symmetry \(f(-x;b,-r)=f(x;b,r)\) allows us to take \(r\ge0\). For \(0<b<3\) the potential tends to \(+\infty\) at both ends of the field range, since the leading behaviour at large \(|x|\) is \(\tfrac{3-b}{8}\ee^{2|x|}\). Let \(k=\sqrt{b/3}\) and define the critical coupling
\begin{equation}
    \rc(b)=\min_{x>0}\frac{\cosh x-k\sinh x}{x},
    \label{eq:rcrit}
\end{equation}
which is attained at the unique positive root \(x_c\) of \(\cosh x\,(1+kx)=\sinh x\,(x+k)\). The following statements are proved in Appendix~\ref{app:vacuum}. For \(0\le r<\rc\) the potential has a unique stationary point \(x_\star\), which is a nondegenerate global minimum with \(f_\star=f(x_\star)>0\). The height \(f_\star\) decreases strictly with \(r\), vanishes at \(r=\rc\) and is negative for \(r>\rc\). The classical dynamics in the positive-vacuum domain is simple. The scalar energy obeys \(\dot\rho_\phi=-3H\dot\phi^2\), and an expanding solution has \(H\ge H_\star\) with \(H_\star^2=\lambda_1A^2f_\star/3\) because \(3H^2=\rho_\phi\ge\lambda_1A^2f_\star\). Decreasing energy together with the coercivity of the potential confines the field and its velocity, and the invariant subset of \(\dot\rho_\phi=0\) is the minimum, so classical solutions approach the de Sitter point \((\phi_\star,\dot\phi=0,H_\star)\). This is a statement about classical solutions and does not imply that all quantum states approach a vacuum. Defining the effective scalar mass at the potential minimum by \(\mu_*^2 \equiv U_{\phi\phi}(\phi_*)\) and the dimensionless mass-to-Hubble ratio by \(\kappa \equiv \mu_*/H_*\), linearization of the Friedmann–Klein–Gordon system around the de Sitter equilibrium gives the characteristic roots
\[
    \lambda_\pm=\frac{H_*}{2}\left(-3\pm\sqrt{9-4\kappa^2}\right),\,
\]
which are real and negative for \(\kappa<3/2\) and form a complex pair with real part \(-3H_\star/2\) for \(\kappa>3/2\), while the perturbation of \(H\) is slaved to the scalar energy and decays with it. The de Sitter point is therefore a stable node for \(\kappa<3/2\) and a stable focus for \(\kappa>3/2\), and the focus makes \(\omega_\star'/(2\pi)\) turns per e-fold of expansion.

Table~\ref{tab:rcrit} gives the critical coupling and the location of the critical minimum for several values of \(b\). The critical coupling decreases with \(b\), since a larger \(b\) makes the negative term \(-\tfrac b2\sinh^2x\) stronger and removes the vacuum for smaller deformations. The case \(b=3/2\) is used for all numerical results below.
%%%%%%%%%%%%%%%%%%%%%%%%%%%%%%%%%%%%%%%%%%%%%%%%%%%%%%%%%%%%%%%%%%%%
\begin{table}[t]
\centering
\caption{Critical coupling \(r_{\rm crit}(b)\), location \(x_c\) of the critical minimum, curvature \(f''(x_c)\) and the undeformed value \(\kappa_0^2=2b(3-b)\).}
\label{tab:rcrit}
\begin{tabular}{ccccc}
\toprule
\(b\) & \(r_{\rm crit}\) & \(x_c\) & \(f''(x_c)\) & \(\kappa_0^2\)\\
\midrule
0.5 & 0.980503 & 1.36631 & 3.00735 & 2.500\\
1.0 & 0.748561 & 1.47978 & 3.99480 & 4.000\\
1.5 & 0.561030 & 1.60889 & 4.59277 & 4.500\\
2.0 & 0.391708 & 1.78114 & 4.92887 & 4.000\\
2.5 & 0.225087 & 2.07115 & 4.98418 & 2.500\\
\bottomrule
\end{tabular}
\end{table}
%%%%%%%%%%%%%%%%%%%%%%%%%%%%%%%%%%%%%%%%%%%%%%%
The local oscillator at the minimum is characterized by \(\mu_\star^2=\lambda_1A^2bf_{xx}(x_\star)\), and the ratio that decides whether the oscillation can mimic dust is
\begin{equation}
    \kappa = \frac{\mu_\star}{H_\star}=\sqrt{\frac{3b\,f_{xx}(x_\star)}{f_\star}} .
    \label{eq:kappa}
\end{equation}

A homogeneous perturbation \(\delta\phi=\phi-\phi_\star\) obeys \(\delta\ddot\phi+3H_\star\delta\dot\phi+\mu_\star^2\delta\phi=0\). It oscillates when \(\kappa>3/2\), with frequency \(\omega_\star=\sqrt{\kappa^2-9/4}\,H_\star\); it is overdamped for \(\kappa<3/2\), and is critically damped at \(\kappa=3/2\). A pressureless description additionally requires \(\kappa\gg1\), so that the envelope changes little over a period. The dependence of \(\kappa\) on the parameters follows from three regimes. At \(r=0\) one finds \(f_\star=3/2\) and \(f_{xx}(0)=3-b\), so \(\kappa_0^2\equiv\kappa^2|_{r=0}=2b(3-b)\), whose maximum over \(0<b<3\) is \(9/2\) at \(b=3/2\). The undeformed hyperbolic potential therefore never produces scale separation, and the values of \(\kappa_0^2\) in Table~\ref{tab:rcrit} are all of order unity. For the baseline potential \(U_{\rm B}=\tfrac\Lambda2(\cosh^2c\phi+1)\) of Ref.~\cite{Bqc1} one has \(U_{{\rm B},\phi\phi}=\Lambda c^2\) and \(H_\star^2=\Lambda/3\) at the minimum, hence \(\kappa^2=3c^2\), which equals \(9/8\) for the value \(c^2=3/8\) required by the canonical transformation and lies below \(9/4\). At small \(r\) the minimum shifts to \(\phi_\star=3\alpha_{\rm c}/[AB^2(3-B^2)]\) and the leading changes are
\begin{equation}
    f_\star=\tfrac32-\frac{9r^2}{2(3-b)}+O(r^4),\qquad f_{xx}(x_\star)=(3-b)-\frac{3b\,r^2}{3-b}+O(r^4),\qquad f_{xxx}(x_\star)=3r+O(r^3),
    \label{eq:small-r}
\end{equation}
so the deformation lowers the vacuum energy, introduces an asymmetric cubic coupling and leaves \(\kappa\) of order unity (Appendix~\ref{app:small}). For \(r=0.02\) and \(b=3/2\) the expansion gives \(f_\star=1.4988\) and agrees with the exact value \(1.49880\). Equation~\eqref{eq:small-r} cannot be extrapolated to large \(r\). The expansion parameter of the shift is \(|B\phi_\star|=3r/(3-b)\) and the formulas become unreliable well before the critical coupling.

\begin{figure}[t]
\centering
\includegraphics[width=\linewidth]{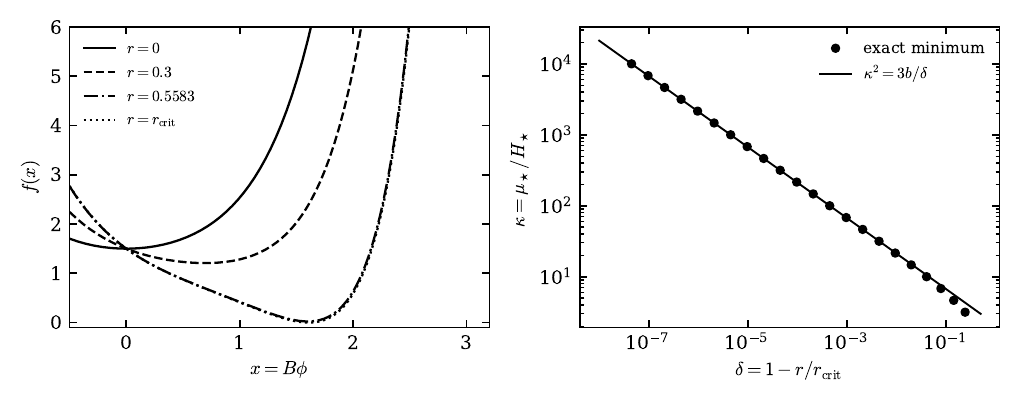}
\caption{Left: the potential \(f(x)\) for \(b=3/2\) and increasing cuscuton coupling \(r\). The minimum moves to larger \(x\) and its height falls to zero at \(r=\rc\). Right: the ratio \(\kappa=\mu_\star/H_\star\) at the exact minimum as a function of \(\delta=1-r/\rc\) (points) compared with the asymptotic prediction \(\kappa\simeq\sqrt{3b/\delta}\) (line).}
\label{fig:potential}
\end{figure}

The opposite limit \(r\to\rc\) is controlled by the proof in the appendix. With \(\delta=1-r/\rc\),
\begin{equation}
    f_\star=f_{xx}(x_c)\,\delta+O(\delta^2),\qquad f_{xx}(x_\star)=f_{xx}(x_c)+O(\delta),\qquad \kappa^2=\frac{3b}{\delta}+O(1).
    \label{eq:kappa-delta}
\end{equation}
For \(b=3/2\) the \(O(1)\) term is a constant close to \(-11.41\), since \(\kappa^2-3b/\delta\) equals \(-11.05\), \(-11.37\), \(-11.40\) and \(-11.41\) at \(\kappa=10,30,100\) and \(1000\). At fixed \(b\) and \(\lambda_1A^2\), the mass scale \(\mu_\star\) stays finite as \(r\to\rc\) while the vacuum energy goes to zero, so the large ratio is produced entirely by the approach to the critical coupling and not by a change in the mass of the field. This is the central structural observation of the paper. The same cuscuton coupling that lowers the vacuum energy also controls the hierarchy between the oscillation frequency and the expansion rate, and the single parameter \(\delta\) fixes the number of oscillations per Hubble time. Figure~\ref{fig:potential} shows the potential for several couplings and compares Eq.~\eqref{eq:kappa-delta} with the exact minimum for \(b=3/2\), for which \(\rc=0.5610303971\) and \(x_c=1.608893\). The two couplings used in the numerical work below are \(r=0.02\), with \(x_\star=0.040021\), \(f_\star=1.49880\) and \(\kappa=2.121321\), and \(r=0.5582602305\), with \(x_\star=1.600449\), \(f_\star=0.0225447\) and \(\kappa=30.0000\), for which \(\delta=4.94\times10^{-3}\).

The anharmonicity of the potential at the minimum does not grow in the same limit. The ratio \(f_{xxx}/f_{xx}\) at the minimum equals \(3.87\) both at \(r=0.5583\) and at \(r=\rc\), so the dimensionless cubic parameter is \(3.87\,y\) for a field excursion \(y\) around the minimum. At fixed excitation energy the amplitude of the oscillation scales as \(y\sim|v|/\kappa\), and the anharmonic corrections to the harmonic picture therefore decrease as \(1/\kappa\). This point is used in Sec.~\ref{sec:observables}.

Equation~\eqref{eq:kappa-delta} also fixes the tuning that a realistic dark matter mass would require. With \(H_\star\simeq1.3\times10^{-33}\) eV and \(\mu_\star=10^{-22}\) eV, a value in the range usually discussed for ultralight dark matter \cite{Hui2017,Irsic2017,Rogers2021}, one needs \(\kappa\simeq8\times10^{10}\) and therefore \(\delta\simeq8\times10^{-22}\). The smallness of the cosmological constant relative to the matter mass is thus converted into the proximity of the cuscuton coupling to its critical value. This reorganizes the hierarchy without providing a mechanism that protects the small value of \(\delta\); the reorganization is traceable to one parameter. The value of \(\delta\) for \(\kappa=30\) is the one used for numerical work because the cell action required to resolve the quantum effects grows with \(\kappa\).

%%%%%%%%%%%%%%%%%%%%%%%%%%%%%%%%%%%%%%%%%%%%%%%%%%%%%%%%%%%%%%%%%%%%%%%%%%%%%%%%%%%%%%%%%%%%%%%%%%%%%%%%%%%%%%%%%%%%%
\section{Adiabatic Gaussian solution and the quantum dust law}
\label{sec:adiabatic}

The structure of the dark sector in the quantum model is most transparent in the regime where the oscillation of the scalar field is fast compared with the expansion and its excitation energy is small. In this regime the leading adiabatic reduction of Eq.~\eqref{eq:wdw} can be solved in closed form around the vacuum, and the Bohmian history decomposes into three parts. We write \(y=x-x_\star\) and \(m_0=\nu/c_x^2\), and we separate the vacuum phase by \(\Psi=\ee^{\ii S_0}\ee^{-3u/2}\varphi\) with \(S_0'=-\nu\ee^{3u}\), which corresponds to \(h=1\) in Eq.~\eqref{eq:guidance}. Here \(\varphi(y,u)\) denotes an auxiliary wave function obtained after extracting the rapidly varying vacuum phase and a scale factor dependent prefactor from the Wheeler–DeWitt wave function. It should not be confused with the canonical scalar field \(\phi\).
The equation for \(\varphi\) is exact,
\begin{equation}
    -2\ii\nu\ee^{3u}\varphi_u+\varphi_{uu}-3\varphi_u+\tfrac94\varphi-c_x^2\varphi_{yy}+\nu^2\ee^{6u}\,[F(x_\star+y)-1]\,\varphi=0 .
    \label{eq:phi-exact}
\end{equation}
Assuming that the auxiliary wave function varies on the scalar oscillation timescale, its derivatives scale as \(\varphi_u=O(\kappa\varphi)\) and \(\varphi_{uu}=O(\kappa^2\varphi)\). Relative to the leading term \(2\nu e^{3u}\varphi_u\), the terms \(\varphi_{uu}\), \(3\varphi_u\), and \(9\varphi/4\) are suppressed by factors of order \(\kappa/(\nu e^{3u})\), \(1/(\nu e^{3u})\), and \(1/(\kappa\nu e^{3u})\), respectively. They can therefore be neglected when \(\nu e^{3u}\gg\kappa\gg1\). We drop the suppressed terms, and we replace \(F-1\) by its quadratic part \(\kappa^2y^2/c_x^2\), which follows from \(F_{xx}(x_\star)=2\kappa^2/c_x^2\). The result is a Schr\"odinger equation \(\ii\varphi_u=H_y\varphi\) with
\begin{equation}
    H_y=\frac{p^2}{2m_0\ee^{3u}}+\tfrac12\,m_0\ee^{3u}\kappa^2y^2 ,
    \label{eq:Hy}
\end{equation}
which is a harmonic oscillator whose mass grows exponentially in the evolution parameter. Such a Hamiltonian is of the Caldirola--Kanai type \cite{Kanai1948,Dekker1981} with friction coefficient \(3\), and the corresponding classical equation of motion is the damped oscillator \(y''+3y'+\kappa^2y=0\) that governs the scalar field around the minimum. Exact Gaussian solutions of such time-dependent oscillators are provided by the invariant method of Lewis and Riesenfeld \cite{Lewis1969}. Here they are obtained more directly from the canonical transformation \(q=\ee^{3u/2}y\), \(P=\ee^{-3u/2}p\), generated by \(F_2=\ee^{3u/2}yP\), which removes the explicit time dependence and produces
\begin{equation}
    K=\frac{P^2}{2m_0}+\tfrac12m_0\kappa^2q^2+\tfrac34(qP+Pq).
    \label{eq:K}
\end{equation}
The Hamiltonian~\(K\) is a positive quadratic form when \(\kappa>3/2\). Its classical orbits are ellipses in the \((q,P)\) plane traversed with the frequency \(\omega_\star'=\sqrt{\kappa^2-9/4}\), and in the original variables they correspond to the damped oscillations of the field. In the quantum theory the transformation is implemented by the dilation \(\varphi(y,u)=\ee^{3u/4}\psi(\ee^{3u/2}y,u)\), which preserves the norm, and \(\ii\psi_u=K\psi\) holds.

The ground state of \(K\) is the Gaussian \(\psi_0\propto\exp[-\tfrac12wq^2]\). Inserting it in \(K\psi_0=E_0\psi_0\) and matching the coefficients of \(q^2\) and \(q^0\) gives (Appendix~\ref{app:gaussian})
\begin{equation}
    w=m_0\left(\omega_\star'+\tfrac32\ii\right),\qquad E_0=\tfrac12\omega_\star'.
    \label{eq:w}
\end{equation}
The imaginary part of \(w\) is the term that distinguishes the present problem from a harmonic oscillator with constant mass. In the original variables the packet has the width \(\sigma_y^2=c_x^2\ee^{-3u}/(2\nu\omega_\star')\), which decreases as the universe expands because the effective mass grows. The resulting leading adiabatic ground-state wave function in the original variables is
\begin{equation}
    \Psi_0=\exp\!\left[-\tfrac{\ii\nu}3\ee^{3u}-\ii E_0u-\tfrac34u-\tfrac12w\,\ee^{3u}y^2\right]\mathcal N,
    \label{eq:Psi0}
\end{equation}
with a constant \(\mathcal N\). Two further families of exact solutions of the reduced oscillator are used below. A Gaussian with a different complex width evolves with a width parameter \(\Gamma(u)\) that obeys a Riccati equation, and linearizing it around \(\Gamma=w\) shows that deviations oscillate as \(\ee^{-2\ii\omega_\star'u}\) in the \(q\) frame. A mismatch between the initial width and the ground-state width therefore excites a breathing mode with the frequency \(2\omega_\star'\) whose amplitude is constant in \(q\). Displaced packets \(\psi_0(q-q_c)\ee^{\ii P_c(q-q_c)}\) are also exact solutions when \((q_c,P_c)\) follows a classical orbit of \(K\), and in the original variables the centre follows the damped oscillation of the field. The state is then a coherent state of the homogeneous mode of the scalar field.

The Bohmian trajectories of these states are simple. For the ground state the phase of Eq.~\eqref{eq:Psi0} is \(S=-\tfrac\nu3\ee^{3u}-E_0u-\tfrac34m_0\ee^{3u}y^2\), and Eq.~\eqref{eq:guidance} gives
\begin{equation}
    v=-\tfrac32y_{\rm B},\qquad h=1+\frac{\omega_\star'}{2\nu}\ee^{-3u}+\frac{9\,y_{\rm B}^2}{4c_x^2}.
    \label{eq:ground-h}
\end{equation}

Using \(\dd y_{\rm B}/\dd u=v/h\), the first relation gives \(y_{\rm B}^{\prime}\simeq-3y_{\rm B}/2\) when \(h\simeq1\). Thus \(y_{\rm B}\propto\ee^{-3u/2}\) and the displacement in units of the packet width is constant to the retained order. Within the reduced Schr\"odinger dynamics, the offset \(q_{\rm B}-q_c\) is constant for displaced packets, because the Bohmian velocity in the \(q\) frame coincides with that of the centre, and the guided configuration oscillates rigidly with the packet. The quantum density follows from the amplitude, \(\ln R=-\tfrac34u-\tfrac12m_0\omega_\star'\ee^{3u}y^2+\mathrm{const}\), inserted in Eq.~\eqref{eq:friedmann}. The terms of order \(\nu\ee^{3u}\) in \(R_{uu}-c_x^2R_{xx}\) are \(c_x^2m_0\omega_\star'\ee^{3u}-c_x^2m_0^2\omega_\star'^2\ee^{6u}y^2\), and the remaining terms are of order one for displacements of order the packet width. Hence

\begin{equation}
    Q\simeq\frac{\omega_\star'}{\nu}\,\ee^{-3u}-\frac{\omega_\star'^2}{c_x^2}\,y^2,\qquad
    \Pi_Q\simeq\frac{\omega_\star'^2}{c_x^2}\,y^2 ,
    \label{eq:Q}
\end{equation}

where the pressure follows from Eq.~\eqref{eq:PiQ} with the derivative taken at fixed \(x\). In terms of the position \(y_{\rm B}=\xi\sigma_y\) of the guided configuration these relations read \(Q\simeq(\omega_\star'/\nu)\ee^{-3u}(1-\xi^2/2)\) and \(\Pi_Q \simeq (\omega_\star'/2\nu)\xi^2\ee^{-3u}\). At the centre of the packet the pressure vanishes at leading adiabatic order, while away from the centre the leading pressure is positive and the quantum density is reduced. The same displacement raises the scalar potential energy by \(\kappa^2y^2/c_x^2\) and the kinetic energy by \(9y^2/(4c_x^2)\). Their sum exceeds the leading loss in \(Q\) by \(9y^2/(2c_x^2)\), using \(\omega_\star'^2=\kappa^2-9/4\). Consequently the total \(h\) in Eq.~\eqref{eq:ground-h} contains \(\xi\) only through a correction of relative order \(9\xi^2/(4\omega_\star'^2)\) to the leading quantum Hubble shift. The split of the dark sector into classical scalar and quantum contributions therefore depends on the position of the guided configuration within the packet, while the expansion rate is insensitive to that position at leading order in the large-\(\kappa\) limit.

The physical content of Eq.~\eqref{eq:Q} is clearest in dimensionful form. Restoring units with \(\nu\ee^{3u}=6VH_\star M_{\rm P}^2/\hbar\) and \(\rho_Q=3H_\star^2M_{\rm P}^2\,Q\) one finds, at the centre of the packet,

\begin{equation}
    \rho_Q\simeq\frac{\hbar\,\omega_\star' H_\star}{2V},
    \label{eq:rhoQ}
\end{equation}

This is the zero-point energy of the effective homogeneous oscillator, with frequency \(\omega_\star'H_\star\), divided by the physical cell volume. A formal Gaussian weighting of the Bohmian displacement gives \(\langle\xi^2\rangle=1\) and hence \(\langle\rho_Q\rangle\simeq\hbar\omega_\star'H_\star/(4V)\) at this order, half the centred value. This weighting is illustrative, since no distribution of initial Bohmian configurations has been assumed. Because \(V\propto a^3\), the centred contribution scales as \(a^{-3}\) and its effective pressure vanishes at leading adiabatic order, without averaging over scalar oscillations. These statements refer to the retained Gaussian approximation; higher-order Wheeler--DeWitt corrections are not included.

The classical excitation of the field has a related description in terms of a coherent state. In the reduced oscillator, its mean occupation number \(\bar n\) gives an excitation \(\omega_\star'\bar n\) above the ground-state eigenvalue of \(K\). Its relation to the classical excitation density is obtained from the coherent-state amplitude and the oscillation average of Sec.~\ref{subsec:occupation_correspondence}:
\begin{equation}
    \bar n\simeq
    \frac{\nu\omega_\star'\ee^{3u}}{2\kappa^2}
    \frac{\overline{\rho}_{\rm exc}}{\Lambda_\star}
    \simeq
    \frac{\overline{\rho}_{\rm exc}V}{\hbar\mu_\star},
    \qquad \kappa\gg1.
    \label{eq:nbar}
\end{equation}
The second relation uses \(\omega_\star'\simeq\kappa\); the overbar denotes the oscillation average in the near-de Sitter, quadratic regime. The instantaneous excitation fraction is \(\Omega_{\rm exc}=\rho_{\rm exc}/(3H^2M_{\rm P}^2)\). For a Hubble-sized cell and \(\kappa=30\) with an excitation fraction of order \(0.3\), the occupation estimate is of order \(10^{119}\). For the cell actions used in Sec.~\ref{sec:numerics}, the corresponding occupation estimate is far smaller, and quantum corrections are numerically accessible. The unified dark sector consists of a constant vacuum contribution, a classical scalar excitation that behaves as dust after rapid-oscillation averaging, and a centred zero-point contribution of relative size \(\kappa/\nu\) that scales as dust at leading adiabatic order.

The classical excitation does not follow the dust law at every instant, and the deviation is the source of the oscillatory features in the expansion history. For the near-de Sitter, quadratic solution \(y=Y\ee^{-3u/2}\cos(\omega_\star'u+\delta_0)\), the excitation density in units of \(\Lambda_\star\) is \((y'^2+\kappa^2y^2)/(6b)\), and a short calculation gives
\begin{equation}
    \rho_{\rm exc}=\bar\rho\left[1+\frac{3\,\omega_\star'}{2\kappa^2}\sin\bigl(2\omega_\star'u+2\delta_0\bigr)+\frac{9}{4\kappa^2}\cos\bigl(2\omega_\star'u+2\delta_0\bigr)\right],\qquad\bar\rho\propto\ee^{-3u},
    \label{eq:stair}
\end{equation}
so that, in this near-de Sitter approximation, the relative modulation is \(3/(2\kappa)\) at large \(\kappa\). The modulation has twice the oscillation frequency because the energy of the excitation is lost mainly while the field moves fast, \(\dot\rho_{\rm exc}=-3H\dot\phi^2\), and it is retained while the field turns around. The Hubble rate therefore follows a staircase that is flat at the turning points of the field and steepest at its crossings of the minimum, and the staircase becomes unresolved as \(\kappa\) grows. The condition \(\nu\ee^{3u}\gg\kappa\) for the adiabatic reduction, the condition \(\kappa>3/2\), a small anharmonic parameter \(3.87\,y\ll1\) and a small ratio \(\rho_{\rm exc}/\Lambda_\star\) for the neglect of the Lorentz-like reduction of the quantum density together define the regime in which the solution of this section applies. The role of the last condition can be seen from Eq.~\eqref{eq:friedmann}. For a packet that moves rigidly with velocity \(\dd x_c/\dd u=v/h\) one has \(R_{uu}\simeq(v/h)^2R_{xx}\), and with \(v^2/c_x^2=h^2-F-Q\) the combination entering \(Q\) becomes \(-c_x^2[(F+Q)/h^2]R_{xx}\), so the quantum density is reduced by the factor \((F+Q)/h^2\) relative to its value at rest; this becomes approximately \(1/h^2\) when \(F\simeq1\) and \(|Q|\ll1\). In the evolution started at \(z_{\rm i}=2\) with \(\kappa=30\) and \(\nu=300\) the initial value of \(Q\nu\ee^{3u}/\omega_\star'\) is \(0.091\) while \(F/h^2=0.091\), in agreement with this estimate. The zero-point law of Eq.~\eqref{eq:Q} therefore holds for \(h\to1\), and this estimate gives a smaller quantum density for a rapidly moving packet than for a packet at rest. In the excited states of Sec.~\ref{sec:numerics} the last two conditions are only marginally satisfied, and the numerical results there should be read as tests of the structure derived here and not as a verification of every coefficient.

%%%%%%%%%%%%%%%%%%%%%%%%%%%%%%%%%%%%%%%%%%%%%%%%%%%%%%%%%%%%%%%%%%%%%%%%%%%%%%%%%%%%%%%%%%%%%%%
\section{Full numerical solution of the Wheeler--DeWitt equation}
\label{sec:numerics}

The adiabatic solution of the previous section neglects terms of relative order \(\kappa/\nu\) and replaces the potential by its quadratic part. To test its predictions without these approximations we integrate Eq.~\eqref{eq:wdw} with the exact potential \(F=f/f_\star\). The carrier frequency of the wave function grows as \(\ee^{3u}\), which makes a direct integration in \(u\) expensive, and we therefore use the variables \(s=\nu\ee^{3u}/3\) and \(\chi=\sqrt s\,\Psi\). The equation becomes
\begin{equation}
    \chi_{ss}=\frac{c_x^2}{9s^2}\chi_{xx}-\left(F+\frac{1}{4s^2}\right)\chi,
    \label{eq:chi}
\end{equation}
which is exact. The field coordinate is restricted to a finite interval with homogeneous Dirichlet conditions, and \(\chi\) is expanded in a sine basis with a collocation grid, so that \(\chi_{xx}\) is diagonal and the potential is a multiplication on the grid, with the transforms done by fast sine transforms. The coefficients of \(\chi\) and \(\chi_s\) are advanced together with the Bohmian position and the proper time by the explicit eighth-order Runge--Kutta method DOP853 \cite{Hairer1993}, using \(\dd x_{\rm B}/\dd s=c_x^2\Imag(\chi_x/\chi)/(9s^2h)\) and \(\dd\tau/\dd s=1/(3sh)\). Spectral discretizations of this kind converge geometrically for smooth solutions \cite{Boyd2001}, and the requirement on the number of modes is set by the largest momentum of the packet, which reaches \(k\simeq\nu\ee^{3u}|v|/c_x^2\).

The guided observables are computed from the wave function and its derivatives at the Bohmian position. In terms of \(A=\chi_s/\chi\) and \(B_x=\chi_x/\chi\),
\begin{equation}
    h=-\Imag A,\qquad v=\frac{c_x^2}{3s}\Imag B_x,\qquad \frac{\dd h}{\dd u}=3s\,h_s+\frac vh\,h_x,
    \label{eq:obs}
\end{equation}
with \(h_s=-\Imag(\chi_{ss}/\chi-A^2)\) and \(h_x=-\Imag(\chi_{sx}/\chi-AB_x)\), where \(\chi_{ss}\) is supplied by Eq.~\eqref{eq:chi} and \(\chi_{sx}\) and \(\chi_x\) by the sine series. The last term in the expression for \(\dd h/\dd u\) is essential, since it carries the motion of the guided configuration. This construction gives \(w_{\rm tot}\) and \(q_{\rm dec}\) without differentiating any plotted curve. The quantum density is obtained from the phase as \(Q=h^2-v^2/c_x^2-F\) and is checked against its amplitude form where the amplitude is resolved.

The classical reference is normalized at the present epoch. We require a vacuum fraction \(\Omega_{\rm vac,0}=0.7\) and an excitation fraction \(0.3\), with the excitation divided equally between kinetic and excess potential energy, which corresponds to the phase \(\pi/4\) of the oscillation, and \(H_{0,{\rm cl}}=70\) km s\(^{-1}\) Mpc\(^{-1}\). The present field position \(x_0>x_\star\) and velocity \(v_0\) follow from \(0.7[F(x_0)-1]=0.3\cos^2\theta\) and \(v_0=-\sqrt{6b\cdot0.3}\,\sin\theta/\sqrt{0.7}\) with \(\theta=\pi/4\), and Eq.~\eqref{eq:classical} is integrated backwards to an initial redshift \(z_{\rm i}\). The scale factor interval \(0\le u\le\ln(1+z_{\rm i})\) is then covered by the Wheeler--DeWitt evolution, and the Hubble rate in physical units is \(H=H_\star h\) with \(H_\star=70\sqrt{0.7}\) km s\(^{-1}\) Mpc\(^{-1}\). This fixes the normalization \(\lambda_1A^2=3H_\star^2/f_\star\) and is an illustration, not a fit, and the partition and phase are inputs. The initial wave function is the Gaussian
\begin{equation}
    \Psi_{\rm i}(x)=\mathcal N\exp\!\left[-\frac{(x-x_{\rm i})^2}{4\sigma_{\rm i}^2}+\ii p_{\rm i}(x-x_{\rm i})\right]W(x),\qquad\sigma_{\rm i}^2=\frac{c_x^2}{2\nu\kappa},\quad p_{\rm i}=\frac{\nu v_{\rm i}}{c_x^2},
    \label{eq:Psii}
\end{equation}
with the vacuum width of the auxiliary oscillator, a phase slope that reproduces the classical velocity, a smooth cutoff \(W\) that vanishes at the boundaries of the field interval, and \(x_{\rm B,i}=x_{\rm i}\). The initial derivative \(\Psi_u\) is supplied by the derivative-corrected rule of Appendix~\ref{app:numerics}, and the second-order equation is then integrated without further approximation. The initial data for the runs reported below are listed in Table~\ref{tab:initial}, and the numerical parameters in Table~\ref{tab:params}.
%%%%%%%%%%%%%%%%%%%%%%%%%%%%%%%%%%%%%%%%%%%%%%%%%%%%%%%%%%%%%%%%%%%%%%%%%%%%%%
\begin{table}[t]
\centering
\caption{Initial data from the backward classical integration with the present partition \(0.7/0.3\) at phase \(\pi/4\). The width \(\sigma_{\rm i}\) is for \(\nu=300\) and scales as \(\nu^{-1/2}\).}
\label{tab:initial}
\begin{tabular}{ccccccc}
\toprule
\(r\) & \(\kappa\) & \(z_{\rm i}\) & \(x_{\rm i}\) & \(v_{\rm i}\) & \(h_{{\rm cl},i}\) & \(\sigma_{\rm i}\)\\
\midrule
0.02 & 2.1213 & 2.0 & 0.501098 & 6.221519 & 2.327145 & 0.08409\\
0.55826 & 30.000 & 2.0 & 1.596648 & \(-9.469202\) & 3.311239 & 0.02236\\
0.55826 & 30.000 & 0.5 & 1.526219 & 2.973603 & 1.575915 & 0.02236\\
\bottomrule
\end{tabular}
\end{table}
\begin{table}[t]
\centering
\caption{Numerical parameters. \(N\) is the number of interior modes, \([x_L,x_R]\) the field interval, and rtol and atol the integrator tolerances. The cutoff has transition width \(0.1\) inside the interval.}
\label{tab:params}
\begin{tabular}{lcccc}
\toprule
Run & \(N\) & \([x_L,x_R]\) & rtol & atol\\
\midrule
Ground state, \(\kappa=30\) & 1023 & \(x_\star\pm0.3\) & \(2\times10^{-12}\) & \(2\times10^{-14}\)\\
Excited, \(\kappa=30\), \(z_{\rm i}=0.5\) & 1535, 3071 & \([x_\star-0.5,\,x_\star+0.4]\) & \(2\times10^{-12}\) & \(2\times10^{-14}\)\\
Excited, \(\kappa=30\), \(z_{\rm i}=2\) & 2047 & \([0.9504,\,2.0504]\) & \(2\times10^{-12}\) & \(2\times10^{-14}\)\\
Small coupling, \(z_{\rm i}=2\) & 1023--3071 & \([-0.3216,\,2.0404]\) & \(2\times10^{-10}\) & \(2\times10^{-12}\)\\
\bottomrule
\end{tabular}
\end{table}

\begin{figure}[t]
\centering
\includegraphics[width=0.52\linewidth]{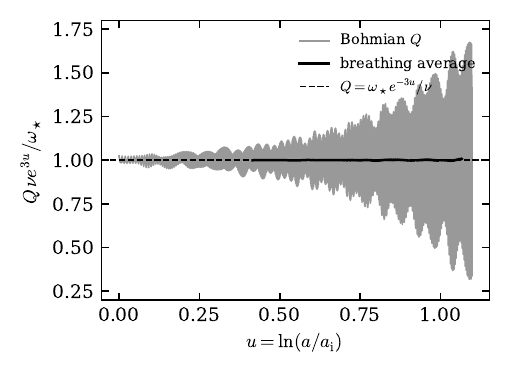}
\caption{Quantum density along the Bohmian trajectory evolved from the reduced-oscillator Gaussian ground state for \(\kappa=30\), \(\nu=300\) and \(\xi=0\), scaled by the prediction \(\omega_\star'\ee^{-3u}/\nu\). The grey curve is the scaled quantum density computed from the full Wheeler--DeWitt evolution, the black curve is the plotted averaged quantity, and the dashed line marks the leading adiabatic value of unity.}
\label{fig:ground}
\end{figure}

The first test isolates the quantum dust law. We start from the Gaussian ground state of the reduced oscillator, Eq.~\eqref{eq:Psi0}, with its analytic \(\Psi_u\), at \(r=0.5582602305\) (\(\kappa=30\)) and integrate over \(0\le u\le\ln3\). The guided configuration is placed at \(\xi\sigma_y\) from the centre. The quantity \(Q\nu\ee^{3u}/\omega_\star'\) oscillates around the leading adiabatic prediction, and we compare its averaged values with the analytic law. The initial analytic data neglect terms of relative order \(\kappa/\nu\); the reduced-oscillator breathing mode has frequency \(2\omega_\star'\), but this alone does not determine all oscillations in the full Wheeler--DeWitt evolution. For \(\xi=0\) and \(\nu=300\), the reported window averages are \(0.9993\), \(0.9998\), \(1.0001\) and \(1.0003\) (Fig.~\ref{fig:ground}). The result for other positions of the guided configuration and for other cell actions is collected in Table~\ref{tab:ground}. The mean of \(Q\) follows \(1-\xi^2/2\) to within \(0.01\) for \(\xi=1\) and \(1.5\), including the sign change at \(\xi=\sqrt2\), and the quantity \(2\nu(h-1)\ee^{3u}/\omega_\star'\) equals one to \(2\times10^{-3}\) for all positions, which is consistent with the weak displacement dependence predicted at large \(\kappa\). The deviation from the prediction decreases with the cell action, from \(6\times10^{-3}\) at \(\nu=100\) to \(3\times10^{-4}\) at \(\nu=300\) and \(10^{-4}\) at \(\nu=1200\).

\begin{figure}[t]
\centering
\includegraphics[width=\linewidth]{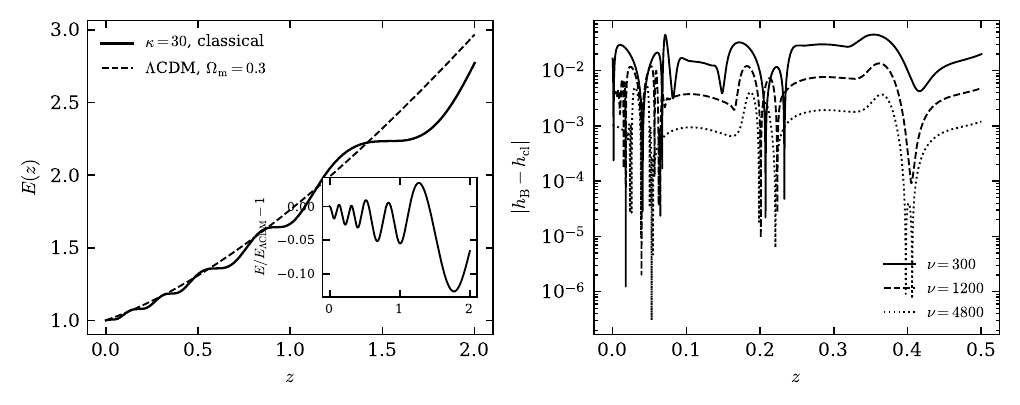}
\caption{Left: classical \(E(z)\) for \(\kappa=30\) from \(z=0\) to \(z=2\), with present vacuum/excitation partition \(0.7/0.3\) and phase \(\theta=\pi/4\), compared with \(\Lambda\)CDM (\(\Omega_{\rm m}=0.3\)); the inset shows \(E/E_{\Lambda{\rm CDM}}-1\). Right: \(|h_{\rm B}-h_{\rm cl}|\) for the Bohmian histories with \(\kappa=30\) started at \(z_{\rm i}=0.5\), for \(\nu=300,1200,4800\).}
\label{fig:history}
\end{figure}

\begin{figure}[t]
\centering
\includegraphics[width=\linewidth]{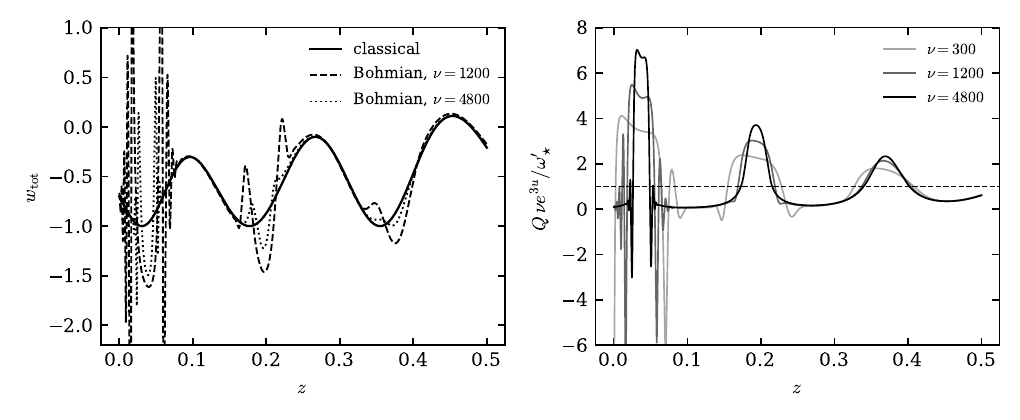}
\caption{Left: total equation of state of the classical history and of the Bohmian histories for \(\nu=1200\) and \(4800\) with \(\kappa=30\) and \(z_{\rm i}=0.5\). Right: scaled quantum density \(Q\nu\ee^{3u}/\omega_\star'\) along the guided trajectory, for \(\nu=300\), \(1200\) and \(4800\). The horizontal dashed line at unity is the centred, near-de Sitter adiabatic benchmark; the excited histories need not follow it pointwise.}
\label{fig:wq}
\end{figure}

\begin{table}[t]
\centering
\caption{%
Test of the zero-point law using the reduced-oscillator Gaussian ground state as initial data for the full Wheeler--DeWitt evolution, \(\kappa=30\). The columns give the displacement \(\xi\) of the guided configuration in units of the packet width, the cell action, the prediction \(1-\xi^2/2\), the reported mean of \(Q\nu\ee^{3u}/\omega_\star'\), and the reported mean of \(2\nu(h-1)\ee^{3u}/\omega_\star'\).}
\label{tab:ground}
\begin{tabular}{ccccc}
\toprule
\(\xi\) & \(\nu\) & \(1-\xi^2/2\) & \(\langle Q\nu\ee^{3u}/\omega_\star'\rangle\) & \(\langle2\nu(h-1)\ee^{3u}/\omega_\star'\rangle\)\\
\midrule
0 & 300 & 1.000 & 1.0003 & 0.9990\\
1.0 & 300 & 0.500 & 0.5087 & 0.9978\\
1.5 & 300 & \(-0.125\) & \(-0.1169\) & 0.9981\\
0 & 100 & 1.000 & 1.0064 & 0.9986\\
0 & 1200 & 1.000 & 0.9999 & 0.9997\\
\bottomrule
\end{tabular}
\end{table}
The second test is an excited condensate with \(\kappa=30\). We start from the reference state at \(z_{\rm i}=0.5\), where the excitation energy is comparable to \(\Lambda_\star\), and evolve to \(z=0\). A repetition of the \(\nu=1200\) run with 2047 modes and tolerances \(2\times10^{-13}\) and \(2\times10^{-15}\) changes \(v\) and \(Q\) along the trajectory by less than \(2\times10^{-10}\) and \(\dd h/\dd u\) by less than \(10^{-8}\), so the numbers below are not limited by discretization. Table~\ref{tab:kappa30} and Figs.~\ref{fig:history} and~\ref{fig:wq} summarize the results for three cell actions, where \(E(z)\equiv H(z)/H_0 = h(z)/h(0)\). The present Hubble rate approaches the classical value \(70\) with a deviation that scales as \(1/\nu\) for \(\nu\ge1200\), namely \(290/\nu\) and \(296/\nu\) in km s\(^{-1}\) Mpc\(^{-1}\), while the maximum of \(|h-h_{\rm cl}|\) over the history decreases more slowly, from \(1.35\times10^{-2}\) to \(5.4\times10^{-3}\) when \(\nu\) grows by a factor four. A rough estimate from the zero-point law, \(\Delta H_0\simeq\tfrac12\Omega_{Q,0}H_{0,{\rm cl}}\) with \(\Omega_{Q,0}=\omega_\star'/(\nu\ee^{3u_0}h_0^2)\) and \(\ee^{3u_0}=(1+z_{\rm i})^3\), gives about \(220/\nu\), which is of the same order as the measured coefficient. The mean of \(Q\nu\ee^{3u}/\omega_\star'\) along the trajectory moves toward the zero-point value, \(1.272\), \(1.118\) and \(1.046\) for \(\nu=300\), \(1200\) and \(4800\). The squared wave amplitude at the Bohmian position never falls below \(0.11\) of its instantaneous spatial maximum, so the guided configuration stays away from nodes. The oscillatory component of the quantum pressure is large at these cell actions and decreases slowly. The total equation of state lies in \([-1.8,0.49]\) at \(\nu=4800\) against the classical range \([-1,0.11]\) and the quantum density oscillates around its mean with swings of several times that mean (Fig.~\ref{fig:wq}). This makes \(w_{\rm tot}\) a more demanding observable than \(h\), and it is the reason why \(h\) and the distances derived from it are used in Sec.~\ref{sec:observables}.
\begin{table}[t]
\centering
\caption{Excited condensate for \(\kappa=30\) starting at \(z_{\rm i}=0.5\). \(H_0\) is in km s\(^{-1}\) Mpc\(^{-1}\) with classical value \(70\). The last column is the minimum over the history of \(|\chi(u,x_{\rm B}(u))|^2/\max_x|\chi(u,x)|^2\).}
\label{tab:kappa30}
\begin{tabular}{rcccc}
\toprule
\(\nu\) & \(H_0\) & \(\max|h-h_{\rm cl}|\) & \(\langle Q\nu\ee^{3u}/\omega_\star'\rangle\) & \(\min|\chi|^2/\max|\chi|^2\)\\
\midrule
300 & 69.031 & \(4.5\times10^{-2}\) & 1.272 & 0.160\\
1200 & 70.242 & \(1.35\times10^{-2}\) & 1.118 & 0.140\\
4800 & 70.062 & \(5.4\times10^{-3}\) & 1.046 & 0.110\\
\bottomrule
\end{tabular}
\end{table}

The third calculation concerns the small coupling \(r=0.02\), for which \(\kappa=2.12\) and the oscillation is not separated from the expansion. We start from \(z_{\rm i}=2\) and list in Table~\ref{tab:small} the quantum corrections for \(100\le\nu\le450\). All entries multiplied by \(\nu\) are constant to better than \(5\%\), and the corrections to the Hubble rate and the total equation of state are bounded by \(0.153/\nu\) and \(0.221/\nu\), respectively, in the tabulated runs. The quantum fraction \(\Omega_Q=Q/h^2\) lies between \(0.078/\nu\) and \(0.204/\nu\) in the tabulated runs. At \(\nu=300\) the largest deviation of the Hubble rate from its classical value is \(5.07\times10^{-4}\) in units of \(H_\star\), so for this coupling the quantum model follows the classical history closely. The classical \(H(z)\) for this coupling has a plateau and deviates from \(\Lambda\)CDM by tens of percent in \(D_V\) below \(z=2\), which shows that the quantum corrections are a finite-\(\nu\) effect on a history that does not resemble that of a dark matter plus dark energy universe. The number of modes matters. With \(N=1023\) at \(\nu=300\) the guided momentum reaches \(k\simeq1250\) near \(z=0\), close to the grid limit \(k_{\max}=N\pi/L_x\simeq1360\), and the Hubble rate acquires oscillatory errors of order \(0.03\) km s\(^{-1}\) Mpc\(^{-1}\) in the last stage of the evolution, which disappear for \(N\ge2047\).
\begin{table}[t]
\centering
\caption{Quantum corrections for \(r=0.02\) (\(\kappa=2.12\)). \(H_0\) is in km s\(^{-1}\) Mpc\(^{-1}\); \(\Delta h\) and \(\Delta w_{\rm tot}\) are differences from the classical history. The entries multiplied by \(\nu\) are nearly independent of \(\nu\).}
\label{tab:small}
\begin{tabular}{rccccc}
\toprule
\(\nu\) & \(H_0\) & \(\nu\max|\Delta h|\) & \(\nu\max|\Delta w_{\rm tot}|\) & \(\nu\Omega_Q^{\min}\) & \(\nu\Omega_Q^{\max}\)\\
\midrule
100 & 70.0159 & 0.1516 & 0.2206 & 0.0785 & 0.2028\\
150 & 70.0105 & 0.1518 & 0.2204 & 0.0791 & 0.2029\\
200 & 70.0077 & 0.1519 & 0.2203 & 0.0794 & 0.2030\\
300 & 70.0051 & 0.1520 & 0.2202 & 0.0798 & 0.2030\\
450 & 70.0034 & 0.1521 & 0.2201 & 0.0800 & 0.2031\\
\bottomrule
\end{tabular}
\end{table}

A start at higher excitation behaves differently. For \(\kappa=30\) from \(z_{\rm i}=2\), where \(\rho_{\rm exc}\simeq10\Lambda_\star\) and the reference velocity is \(v_{\rm i}=-9.47\), the same code with \(\nu=300\) gives \(H_0=70.540\) km s\(^{-1}\) Mpc\(^{-1}\), so the present expansion rate again lies within one percent of the classical value. The guided configuration approaches a node of the wave function, with \(|\chi(x_{\rm B})|^2\) falling to \(1.3\times10^{-3}\) of its maximum, and \(w_{\rm tot}\) then shows narrow spikes of modulus up to \(10^4\) because it contains a derivative of the phase. The Hubble rate itself stays regular. For such states the first-derivative quantities are robust, the second-derivative quantities are not controlled by the present resolution, and the results for \(\kappa=30\) therefore rest on the \(z_{\rm i}=0.5\) histories.

%%%%%%%%%%%%%%%%%%%%%%%%%%%%%%%%%%%%%%%%%%%%%%%%%%%%%%%%%%%%%%%%%%%%%%%%%
\section{Late-time observables and the \texorpdfstring{\(\Lambda\)CDM}{LambdaCDM} limit}
\label{sec:observables}

Baryon acoustic oscillation (BAO) measurements constrain distances measured in units of the sound horizon \(r_{\rm d}\) at the drag epoch \cite{Eisenstein2005,DESI}. To characterize the predicted expansion history, we introduce the Hubble distance \(D_H(z)=c/H(z)\), the transverse comoving distance \(D_M(z)=c\int_0^z dz'/H(z')\), and the volume-averaged distance \(D_V(z)=[zD_M^2(z)D_H(z)]^{1/3}\), appropriate for a spatially flat FLRW background. These quantities enter observational BAO constraints through their ratios to the sound horizon \(r_d\). For a spatially flat background the relevant combinations constrained by BAO analysis are:
\begin{equation}
    \frac{D_H}{r_{\rm d}}=\frac{c}{H_0r_{\rm d}}\frac1{E(z)},\qquad\frac{D_M}{r_{\rm d}}=\frac{c}{H_0r_{\rm d}}\int_0^z\frac{\dd z'}{E(z')},\qquad\frac{D_V}{r_{\rm d}}=\left[z\left(\frac{D_M}{r_{\rm d}}\right)^2\frac{D_H}{r_{\rm d}}\right]^{1/3},
    \label{eq:BAO}
\end{equation}
with \(E(z)=H(z)/H_0\). The model studied in this article has no baryon--photon plasma and cannot compute \(r_{\rm d}\), so only the shape of \(E(z)\) can be compared with a reference cosmology, with \(c/(H_0r_{\rm d})\) treated as a free calibration. In the model the effective dark energy is the constant vacuum term, so \(w_{\rm DE}=-1\) exactly, and the distinctive prediction is the form of the departure of \(E(z)\) from a pure dust plus vacuum law. In the near-de Sitter, quadratic approximation, Eq.~\eqref{eq:stair} describes an oscillatory departure with frequency \(2\omega_\star'\) in \(\ln(1+z)\) and relative excitation-energy amplitude of order \(1/\kappa\). Over a more general background, the phase depends on the evolving Hubble rate. The quantum corrections do not contribute at an observable level for physical cells, where \(\kappa/\nu\sim10^{-120}\kappa\), so the observable content is the classical history.

\begin{figure}[t]
\centering
\includegraphics[width=0.52\linewidth]{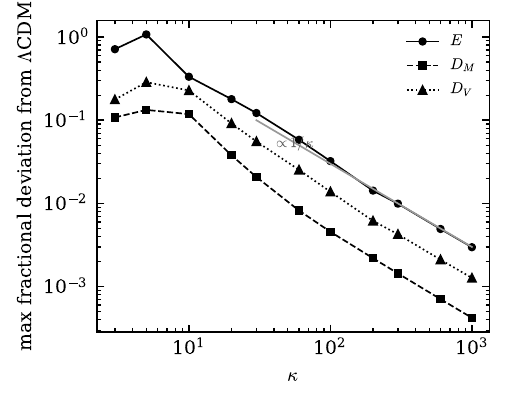}
\caption{Maximal fractional deviation of \(E(z)\), \(D_M\) and \(D_V\) from \(\Lambda\)CDM over \(0.05<z\le2\) as a function of \(\kappa\), for the classical reference with present partition \(0.7/0.3\) (median over twelve oscillation phases). The grey line shows \(3/\kappa\).}
\label{fig:kappa}
\end{figure}

We quantify the approach to \(\Lambda\)CDM using the classical reference and a family of couplings. For each target \(\kappa\) we choose \(r\) so that the exact minimum has that value of \(\kappa\), normalize the reference at \(z=0\) to vacuum and excitation fractions \(0.7\) and \(0.3\) at twelve equally spaced phases of the oscillation, integrate to \(z=2\), and compare \(E\), \(D_M\) and \(D_V\) with \(\Lambda\)CDM with \(\Omega_{\rm m}=0.3\). Table~\ref{tab:kscan} and Fig.~\ref{fig:kappa} give the median over phases of the maximal fractional deviation for \(0.05<z\le2\). All three measures decrease as \(1/\kappa\) at large \(\kappa\). For \(\kappa\ge100\) the deviations approach \(3.0/\kappa\) for \(E\), \(0.42/\kappa\) for \(D_M\) and \(1.3/\kappa\) for \(D_V\). The near-de Sitter expression in Eq.~\eqref{eq:stair} illustrates how rapid scalar oscillations produce corrections of order \(1/\kappa\). The coefficients reported here are empirical results for the specified background normalization, redshift interval and phase sampling. They also include the difference between the instantaneous present partition and the oscillation-averaged partition, together with anharmonic corrections. At fixed dimensionless excitation energy, the local anharmonic parameter \(3.87\,y\) decreases as \(1/\kappa\), as discussed in Sec.~\ref{sec:potential}. Distances are much less sensitive than \(H\) because the oscillation averages out in the integral, and the deviation of \(D_M\) is a factor of seven smaller than that of \(E\). For rapid oscillations in the quadratic regime, the number of scalar oscillation periods over \(0\le z\le2\) is approximately \(N_{\rm osc}\simeq[\kappa/(2\pi)]\int_0^2\dd z/[(1+z)h(z)]\), rather than \(\omega_\star'\ln3/(2\pi)\) unless \(h\simeq1\) throughout. The leading energy modulation has twice the scalar phase. Counts for the full potential should be obtained directly from the numerical histories.

\begin{table}[t]
\centering
\caption{Classical reference: median over twelve oscillation phases of the maximal fractional deviation from \(\Lambda\)CDM (\(\Omega_{\rm m}=0.3\)) over \(0.05<z\le2\), and the maximum over phases for \(D_V\).}
\label{tab:kscan}
\begin{tabular}{rccccc}
\toprule
\(\kappa\) & \(\delta\) & \(E\) & \(D_M\) & \(D_V\) & \(D_V\) (max over phase)\\
\midrule
5 & 0.128 & 1.07 & 0.133 & 0.287 & 0.404\\
10 & 0.0405 & 0.333 & 0.118 & 0.228 & 0.307\\
20 & 0.0109 & 0.179 & 0.0379 & 0.0916 & 0.114\\
30 & 0.0049 & 0.122 & 0.0208 & 0.0556 & 0.0625\\
60 & 0.0012 & 0.0582 & 0.0082 & 0.0253 & 0.0282\\
100 & \(4.5\times10^{-4}\) & 0.0323 & 0.0046 & 0.0138 & 0.0164\\
200 & \(1.1\times10^{-4}\) & 0.0143 & 0.0022 & 0.0062 & 0.0071\\
300 & \(5.0\times10^{-5}\) & 0.0099 & 0.0014 & 0.0043 & 0.0049\\
600 & \(1.25\times10^{-5}\) & 0.0049 & 0.0007 & 0.0021 & 0.0024\\
1000 & \(4.5\times10^{-6}\) & 0.0030 & 0.0004 & 0.0013 & 0.0014\\
\bottomrule
\end{tabular}
\end{table}

The departure from \(\Lambda\)CDM contains an oscillatory contribution from the scalar excitation. Equation~\eqref{eq:stair} describes this contribution in the near-de Sitter, quadratic regime, where \(H\simeq H_\star\). More generally, writing \(y=x-x_\star\) and using primes for derivatives with respect to \(u\), the classical scalar equation in the quadratic approximation is
\begin{equation}
    y''+\left(3+\frac{h'}{h}\right)y'
    +\frac{\kappa^2}{h^2}y=0.
    \label{eq:review_scalar_u_general}
\end{equation}
For rapid oscillations, \(\mu_\star\gg H\), the leading scalar phase is
\begin{equation}
    \Theta(z)\simeq\Theta_0-\kappa\int_0^z
    \frac{\dd z'}{(1+z')h(z')}.
    \label{eq:review_phase_redshift}
\end{equation}
The leading oscillatory contribution to the excitation energy has twice the scalar phase. A constant period \(\pi/\omega_\star'\) in \(\ln(1+z)\) follows in the near-de Sitter approximation; it is not a general property of the entire redshift interval considered here. Outside that limit, the phase and amplitude must be obtained from the evolving background. Oscillatory departures from a dust law arise generically for oscillating scalar fields \cite{Turner1983}. The distinctive feature of the present potential is the relation between the vacuum mass-to-Hubble ratio \(\kappa\) and the distance \(\delta\) from criticality.

Using the empirical large-\(\kappa\) scaling reported in Table~\ref{tab:kscan}, a median maximal fractional difference below one percent in \(D_V\) relative to the chosen \(\Lambda\)CDM reference corresponds approximately to \(\kappa\gtrsim130\), or \(\delta\lesssim3\times10^{-4}\). A half-percent criterion corresponds approximately to \(\kappa\gtrsim260\). These are illustrative similarity thresholds for the stated normalization, redshift interval and phase sampling; they are not observational confidence limits from a DESI likelihood analysis. The model contains no mechanism to lift \(w\) away from \(-1\) in the dark energy sector, and it cannot reproduce the time-dependent dark energy that is preferred at the level of a few standard deviations when DESI baryon acoustic oscillation data are combined with supernova and microwave background data \cite{DESI2024,DESI}. If that preference is confirmed, the vacuum component of the present model would have to be generalized, for instance by allowing the field to be slowly rolling, and the identification of the cuscuton coupling as the controller of the hierarchy would carry over to that setting.

The condensate that makes up the dark matter also has perturbations, which are not described by the homogeneous model. For an oscillating scalar field with mass \(\mu_\star\) the relevant scale is the quantum Jeans wavenumber \(k_J\simeq a(16\pi G\rho)^{1/4}\mu_\star^{1/2}\) \cite{Hu2000,Johnson2008}, which in units of \(aH\) is of order \(\kappa^{1/2}\) times a number of order one. Fluctuations with smaller wavelengths are suppressed. For \(\kappa\) of order tens the Jeans wavenumber is a few times \(aH\), corresponding to an inverse-wavenumber length smaller than the Hubble radius, and the model as a literal description of dark matter requires \(\kappa\) to be large enough that \(k_J\) lies at comoving scales of galaxies, which is the range explored by studies of ultralight dark matter with the Lyman-\(\alpha\) forest and the microwave background \cite{Hlozek2015,Irsic2017,Rogers2021}. The background results are independent of this requirement, and we regard the scale \(\kappa\sim10^{10}\) to \(10^{11}\) as the one at which the model would have to be confronted with structure formation.
%%%%%%%%%%%%%%%%%%%%%%%%%%%%%%%%%%%%%%%%%%%%%%%%%%%%%%%%%%%%%%%%%%%%%%%%%%%%%%%%
\section{Semiclassical limit of Bohmian Wheeler--DeWitt cosmology}
\label{sec:semiclassical_limit}

The numerical results presented in the preceding sections demonstrate that the Bohmian cosmological histories approach their classical counterparts as the minisuperspace action parameter \(\nu\) increases. This provides a controlled semiclassical limit of the Wheeler--DeWitt (WDW) formulation, in which the homogeneous geometry and scalar field follow trajectories approaching the classical Einstein--Klein--Gordon evolution.

The dimensionless WDW equation: \(\Psi_{uu}-c_x^2\Psi_{xx}+\nu^2e^{6u}F(x)\Psi=0\), is written in terms of variables and parameters as: \(\nu=6V_iH_\star\), \(u=\ln(a/a_i)\), \(x=B\phi\),  \(c_x^2=6b\) and \(F(x)=\frac{f(x)}{f_\star}\). Here \(V_i=V_0a_i^3\) is the physical volume of the homogeneous cell at the reference epoch and \(H_\star^2=\Lambda_\star/3\). We use units in which \(\hbar=1\). Along with the Friedmann constraint \(h^2=K+F+Q\), where \(K=\frac{v^2}{c_x^2}\), we can rewrite the relevant Bohmian evolution equations as:
\begin{align}
    \frac{dx_B}{du}&=\frac{v_B}{h_B},\\
    \frac{dv_B}{du}&=-3v_B-\frac{3b}{h_B}(F_x+Q_x),\\
    \frac{dh_B}{du}&=-\frac{3K_B}{h_B}
    +\frac{(\partial_uQ)_x}{2h_B}.
    \label{eq:sc_bohm_evolution}
\end{align}
All quantities on the right-hand sides are evaluated along the Bohmian trajectory. The subscript on \((\partial_u Q)_x\) emphasizes that the partial derivative is taken at fixed \(x\), rather than along the trajectory. When \(Q\) and its dynamically relevant derivatives become negligible, these equations reduce to
\begin{align}
    h_{\rm cl}^2&=K_{\rm cl}+F(x_{\rm cl}),\\
    \frac{dx_{\rm cl}}{du}&=\frac{v_{\rm cl}}{h_{\rm cl}},\\
    \frac{dv_{\rm cl}}{du}&=-3v_{\rm cl}
    -\frac{3b}{h_{\rm cl}}F_x(x_{\rm cl}),\\
    \frac{dh_{\rm cl}}{du}&=-\frac{3K_{\rm cl}}{h_{\rm cl}}.
    \label{eq:sc_classical_system}
\end{align}
These are the classical Friedmann, Klein--Gordon, and Raychaudhuri equations in the variables of the present model.

The conventional WKB correspondence can be exhibited by writing
\begin{equation}
    \Psi(u,x)={\mathcal R}_0(u,x)\exp\!\left[i\nu S_0(u,x)\right],
    \label{eq:sc_wkb}
\end{equation}
where \(S_0\) and \(\mathcal R_0\) denote the leading-order phase and amplitude, respectively, in the asymptotic WKB expansion in inverse powers of the dimensionless minisuperspace action parameter \(\nu\). The leading order of the WDW equation yields:
\begin{equation}
    (S_{0,u})^2-c_x^2(S_{0,x})^2=e^{6u}F(x),
    \label{eq:sc_wkb_hj}
\end{equation}
which is the classical minisuperspace Hamilton--Jacobi equation. At the next order, the amplitude satisfies the transport equation
\begin{equation}
    2S_{0,u}{\mathcal R}_{0,u}-2c_x^2S_{0,x}{\mathcal R}_{0,x}
    +{\mathcal R}_0\left(S_{0,uu}-c_x^2S_{0,xx}\right)=0.
    \label{eq:sc_transport}
\end{equation}
Consequently, a single smooth WKB branch gives classical Bohmian dynamics at leading order, although the WDW state remains a quantum state. For an amplitude with \(\nu\)-independent logarithmic derivatives, \(Q=O(\nu^{-2})\); the Gaussian families used here have width proportional to \(\nu^{-1/2}\), so their transverse amplitude derivatives instead allow the observed \(Q=O(\nu^{-1})\) scaling.

The full-potential calculations of Sec.~\ref{sec:numerics} support the emergence of the classical limit beyond the local Gaussian approximation. For the small-deformation case \(r=0.02\), the maximum difference between the Bohmian and classical dimensionless Hubble rates decreases approximately as \(0.152/\nu\) over the investigated interval. For the heavier-field case \(\kappa=30\), the present-day Hubble correction also exhibits approximately inverse-\(\nu\) scaling at sufficiently large cell actions, although the maximum deviation over the full evolution converges more slowly. These results provide numerical evidence that the Bohmian expansion histories approach their classical counterparts as the minisuperspace action parameter increases. In the small-deformation example \(r=0.02\), the quantum fraction
\begin{equation}
    \Omega_Q=\frac{Q}{h^2}
    \label{eq:sc_omega_q}
\end{equation}
also exhibits an approximately inverse-\(\nu\) dependence. These numerical scalings characterize the state families examined here, not arbitrary WDW solutions. In particular, derivative-sensitive observables can converge more slowly, and configurations passing near nodes require additional care.

For the centred, near-de Sitter adiabatic Gaussian state of Sec.~\ref{sec:adiabatic}, the leading quantum contribution takes the explicit form
\begin{equation}
    Q(u,x_B)\simeq\frac{\omega_\star'}{\nu}e^{-3u},\qquad
    \omega_\star'=\sqrt{\kappa^2-\frac94},\qquad
    \kappa=\frac{\mu_\star}{H_\star},
    \label{eq:sc_gaussian_q}
\end{equation}
under the conditions \(\nu e^{3u}\gg\kappa\gg1\). Hence
\begin{equation}
    \rho_Q=\Lambda_\star Q
    \simeq\frac{\omega_\star'H_\star}{2V_i}e^{-3u}
    =\frac{\omega_\star'H_\star}{2V(u)},
    \qquad V(u)=V_i e^{3u}.
    \label{eq:sc_quantum_density}
\end{equation}
The centered-state quantum contribution therefore scales as \(a^{-3}\) without averaging over classical field oscillations, and its leading effective quantum pressure vanishes:
\begin{equation}
    \Pi_Q=-Q-\frac13(\partial_u Q)_x\simeq0.
    \label{eq:sc_pressure}
\end{equation}
The corresponding fractional quantum term is
\begin{equation}
    \Omega_Q\simeq\frac{\omega_\star'}{\nu h^2}e^{-3u}.
    \label{eq:sc_gaussian_fraction}
\end{equation}
For fixed \(\kappa\), this is suppressed as \(\nu\) increases. In a vacuum-dominated epoch with \(h\) approaching a nonzero constant, it also decreases during expansion. Nevertheless, a small \(\Omega_Q\) alone is not sufficient to establish negligible quantum dynamics: the derivatives \(Q_x\) and \((\partial_u Q)_x\) must also be controlled.

Two distinct physical limits arise in our model. The large-\(\kappa\) limit separates the classical scalar oscillation timescale from the Hubble timescale and permits an approximately pressureless \emph{oscillation-averaged} classical excitation. By contrast, the large-\(\nu\) limit suppresses quantum corrections for the WDW state families considered. Thus the approach toward a dust-plus-vacuum background and the emergence of semiclassical trajectories are logically different phenomena. The late-time \(\Lambda\)CDM comparison of Sec.~\ref{sec:observables} is primarily controlled by the classical scalar dynamics, while the Bohmian analysis quantifies the additional state-dependent correction and its suppression.

%%%%%%%%%%%%%%%%%%%%%%%%%%%%%%%%%%%%%%%%%%%%%%%%%%%%%%%%%%%
\subsection{Occupation-number correspondence between quantum dust and the classical scalar condensate}
\label{subsec:occupation_correspondence}

From Sec.~\ref{sec:adiabatic} we know that we can work with the Hamiltonian \(K\), which is related by a time-dependent canonical transformation to the auxiliary Hamiltonian \(H_y\) and \(K
= \frac{P^{\,2}}{2m_0} + \frac12m_0\kappa^2q^2 + \frac34 \left(q P + Pq\right).\) A further momentum shift, \({\mathcal P} = P+\frac32m_0q,\) brings the Hamiltonian into the standard oscillator form
\begin{equation}
    K =
    \frac{{\mathcal P}^{\,2}}{2m_0}
    +
    \frac12m_0\omega_\star'^2q^2.
    \label{eq:occ_standard_K}
\end{equation}
Defining
\begin{equation}
    a
    =
    \sqrt{\frac{m_0\omega_\star'}{2}}\,q
    +
    \frac{i{\mathcal P}}
    {\sqrt{2m_0\omega_\star'}},
\end{equation}
and its Hermitian conjugate \(a^\dagger\), we obtain
\begin{equation}
    [a,a^\dagger]=1,
    \qquad
    K
    =
    \omega_\star'
    \left(a^\dagger a+\frac12\right).
    \label{eq:occ_number_hamiltonian}
\end{equation}
The eigenvalues of the reduced oscillator Hamiltonian, \(K\), are therefore:
\begin{equation}
    E_n
    =
    \omega_\star'
    \left(n+\frac12\right),
    \qquad
    n=0,1,2,\ldots.
    \label{eq:occ_spectrum}
\end{equation}
Since \(u\) is dimensionless, \(H_\star E_n\) is the corresponding eigenvalue of the cosmic-time evolution generator in the transformed variables when \(H\simeq H_\star\). The physical scalar excitation energy is evaluated from the field and its velocity below.

To connect this quantum description with the classical scalar excitation, consider a coherent state
\(|\alpha\rangle\) satisfying
\begin{equation}
    \widehat a|\alpha\rangle=\alpha|\alpha\rangle,
    \qquad
    \bar n\equiv \langle \alpha|a^\dagger a|\alpha \rangle = |\alpha|^2.
    \label{eq:occ_coherent}
\end{equation}
The expectation value of the oscillator coordinate is
\begin{equation}
    \langle q(u)\rangle
    =
    \sqrt{\frac{2\bar n}{m_0\omega_\star'}}
    \cos(\omega_\star'u+\theta).
    \label{eq:occ_q_expectation}
\end{equation}
Returning to the original field displacement gives
\begin{equation}
    \langle y(u)\rangle
    =
    Y_0 e^{-3u/2}
    \cos \Theta, \,\,\,\,\,\,\,\,Y_0 \equiv \sqrt{\frac{2\bar n}{m_0\omega_\star'}}\,,\,\,\,\,\Theta\equiv \omega_\star'u+\theta\,.
    \label{eq:occ_y_expectation}
\end{equation}
Thus, the center of the coherent state follows the classical damped oscillator equation and its oscillation amplitude decreases as \(a^{-3/2}\). If we write \(y_{\rm cl}\) to be the classical solution of \(H_y\), we know it satisfies \(y^{\prime \prime}_{\rm cl} + 3y^\prime_{\rm cl} + \kappa^2 y_{\rm cl}=0\), where prime stands for a derivative with respect to \(u\). Its solution is
\begin{eqnarray}
    y_{\rm cl}(u)
    = C e^{-3u/2}\cos(\omega_*'u+\delta).
\end{eqnarray}
Choosing \(C=Y_0\) and \(\delta=\theta\) we obtain
\begin{eqnarray}
    \langle y(u) \rangle = y_{\rm cl}(u)\,,
    \label{impl}
\end{eqnarray}
In general for the scalar field, \(\rho_{\rm exc}
=\Lambda_*(K+F-1)\), (where \(K\) is the dimensionless scalar kinetic contribution and not the transformed Hamiltonian) and in the quadratic approximation (of the potential):
\begin{eqnarray}
    \rho_{\rm exc} =
    \frac{H_*^2}{2B^2}
    \left(y'^2+\kappa^2y^2\right),\qquad
    p_{\rm exc}
    =
    \frac{H_*^2}{2B^2}
    \left(y'^2-\kappa^2y^2\right)\,,
    \label{rhop}
\end{eqnarray}
where we have used the approximation \(H\simeq H_*\) near the minimum. In this correspondence the excitation density and pressure are evaluated on the coherent-state centre; fluctuation contributions are treated separately. Now for a generic function \(g(u)\), the averaging over \(u\) is defined as:
\begin{eqnarray}
    \overline{g}(u)
    =
    \frac{1}{T_u}
    \int_u^{u+T_u}g(s)\,ds.
    \label{avrg1}
\end{eqnarray}
where \(T_u=2\pi/\omega^\prime_*\). For rapid oscillations, we treat the envelope \(e^{-3u/2}\) as approximately constant during this interval. In this convention:
\[
    \overline{\sin^2\Theta}=\frac12,
    \qquad
    \overline{\sin\Theta\cos\Theta}=0\,,\qquad
    \overline{e^{-3u}\cos^2\Theta}
    \simeq
    \frac12e^{-3u}.
\]
Using this averaging scheme we get:
\begin{eqnarray}
    \overline{y^2}
    \simeq\frac{Y_0^2}{2}e^{-3u}\,,\quad
    \overline{y'^2}
    \simeq
    \frac{\kappa^2Y_0^2}{2}e^{-3u}\,,
\end{eqnarray}
where, in deriving the second equation, we have used the relation \(\omega_*'^2=\kappa^2-\frac94.\) Using these averages we have
\begin{eqnarray}
    \overline{\rho_{\rm exc}}
    \simeq
    \frac{H_*^2\kappa^2Y_0^2}{2B^2}e^{-3u}
    \propto a^{-3}\,,\qquad
    \overline{p_{\rm exc}}\simeq0.
    \label{prhoav}
\end{eqnarray}
For averaging over many oscillations, a coarse-graining window \(\Delta u\) must satisfy 
\begin{eqnarray}
    \frac{2\pi}{\omega_*'}\ll\Delta u\ll1\,,
\end{eqnarray}
this is possible when \(\omega^\prime_* \gg 1\) or equivalently \(\kappa \gg 1\).

The variance,  \((\Delta q)^2 \equiv \langle q^2 \rangle_\alpha - \langle q \rangle^2_\alpha  =1/(2m_0 \omega^\prime_*)\), is independent of the mean occupation number \(\bar{n}=|\alpha|^2\) and in \(y\) coordinate it looks like:
\[
    (\Delta y)^2
    =
    \frac{e^{-3u}}{2m_0\omega_*'}.
\]
The slowly decreasing amplitude of the oscillations, in Eq.~(\ref{eq:occ_y_expectation}) is \(Y(u) \equiv Y_0 e^{-3u/2}\). To determine whether the quantum state is localized relative to its classical oscillatory motion, we compare the width with the oscillation envelope:
\begin{eqnarray}
    \frac{\Delta y}{Y(u)}
    =
    \frac{1}{2\sqrt{\bar n}}\,,
\end{eqnarray}
showing that for large enough \(\bar{n}\), the solution has a sharp semiclassical correspondence.

One can now show that, in the above limits, the coherent state results produce the correct time average energy density and pressure. That will show that the emergence of a classical fluid like behavior can emerge from a quantum perspective. In the classical solution we had \(\Theta = \omega^\prime_* u + \theta\) and the \(T_u=2\pi/\omega^\prime_*\). Near the potential minimum:
\[
    \frac{du}{dt}=H \simeq H_*\,,
\]
and \(u \simeq H_*t + {\rm constant}\), and consequently we can write \(\Theta = \omega^\prime_* H_* t + \theta_0\). Thus \(T_t=2\pi/(\omega^\prime_*H_*)\) and \(T_u=H_*T_t\). The time average for any generic function \(g(t)\) is given by:
\[
    \overline g_t
    =
    \frac{1}{T_t}
    \int_t^{t+T_t}g(t')\,dt'.
\]
Using \(dt^\prime = du/H_*\) one has:
\begin{eqnarray}
    \overline g_t =
    \frac{1}{H_*T_t}
    \int_u^{u+H_*T_t}g(u')\,du' =\overline{g}_u.
\end{eqnarray}
This equality is exact when \(H=H_*\) is constant and the same corresponding endpoints are used. Using this correspondence one can now convert the previous \(u\)-averages over the energy density and pressure into \(t\)-averages, which naturally appears in classical fluid averaging procedure in cosmology.  In our approximation limits, we intuitively recover the classical results from a quantum state: the large-\(\kappa\) limit justifies the classical time averaging, whereas the large-\(\bar n\) limit justifies interpreting the coherent quantum state as a classical oscillating scalar field. For our Wheeler–DeWitt reduction to the Schr{\"o}dinger oscillator to remain valid, we also require \(\nu e^{3u}\gg\kappa\), together with sufficiently small field excursions for the quadratic potential approximation. Thus, the complete hierarchy relevant to our correspondence is
\begin{eqnarray}
    \nu e^{3u}\gg\kappa\gg1,
    \qquad
    \bar n\gg1,
\end{eqnarray}
subject to the quadratic approximation remaining valid.

In deriving the above classical correspondence we have also assumed that the scalar excitation remains small compared with the vacuum energy:
\[
    \frac{\overline{\rho}_{\rm exc}}{\Lambda_*}\ll1.
\]
Using the coherent-state amplitude, this becomes
\[
    \frac{\overline{\rho}_{\rm exc}}{\Lambda_*}
    \simeq
    \frac{2\kappa^2\bar n}{\nu\omega_*'}e^{-3u}.
\]
This condition gives an additional requirement:
\begin{eqnarray}
    1\ll\bar n
    \ll
    \frac{\nu\omega_*'}{2\kappa^2}e^{3u}.
\end{eqnarray}
For \(\kappa\gg1\) so that \(\omega^\prime_* \simeq \kappa\), we get an upper bound for \(\bar{n}\):
\begin{eqnarray}
    1\ll\bar n
    \ll
    \frac{\nu e^{3u}}{2\kappa}.
\end{eqnarray}
This is an important condition because increasing \(\bar n\) indefinitely at fixed \(\nu,\kappa,u\) eventually invalidates the near-de Sitter approximation.

%%%%%%%%%%%%%%%%%%%%%%%%%%%%%%%%%%%%%%%%%%%%%%%%%%%%%%%%%%%%%%%%%%%%%%%%%%%%%%%%%%%%%
%%%%%%%%%%%%%%%%%%%%%%%%%%%%%%%%%%%%%%%%%%%%%%%%%%%%%%%%%%%%%%%%%%%%%%%%%%%%%%%%%%%%
\section{Discussions and Conclusion}
\label{sec:conclusions}

We have asked what a single canonical scalar field can contribute to the dark sector when its potential is the cuscuton-deformed hyperbolic function and its cosmological state is a solution of the Wheeler--DeWitt equation interpreted in the de Broglie--Bohm framework. The answer has a classical part and a quantum part, and the two are linked by one parameter.

The classical part concerns the vacuum. For \(0<b<3\) the potential has a unique nondegenerate minimum with positive height as long as the cuscuton coupling lies below the critical value \(\rc(b)\) of Eq.~\eqref{eq:rcrit}, and the height vanishes linearly in \(\delta=1-r/\rc\) at the boundary. The curvature at the minimum stays finite in this limit, so the ratio of the scalar mass to the de Sitter Hubble rate obeys \(\kappa^2=3b/\delta+O(1)\). In dimensionful form the vacuum energy density is \(\rho_\Lambda\simeq\mu_\star^2M_{\rm P}^2\,\delta/b\), so the smallness of the cosmological constant relative to the mass scale of the field is exactly the smallness of \(\delta\). The same quantity counts the oscillations of the field per Hubble time, since the oscillation frequency in units of \(H_\star\) is \(\omega_\star'\simeq\kappa\). This is the statement that makes a single scalar a candidate for both components. Without the deformation the hierarchy is absent, since \(\kappa^2=2b(3-b)\le9/2\), and the exactly solvable baseline potential has \(\kappa^2=3c^2=9/8<9/4\), so the scalar is overdamped near the minimum and cannot supply a rapidly oscillating dust-like component in that regime. The deformation therefore does more than lower the vacuum energy. It creates the only regime in which an oscillating condensate exists with a period short compared with the expansion time. The price is a tuning of \(\delta\) to \(\sim10^{-21}\) if the mass is in the ultralight range, and we regard this as the relocation of the cosmological constant problem into the proximity of one coupling to its critical value, with the advantage that the relocation is exact and traceable to a single parameter. The model does not predict the ratio \(\Omega_\Lambda/\Omega_{\rm m}\), which is fixed here by the present partition of the field energy, in the same way as in \(\Lambda\)CDM.

The quantum part concerns what the Wheeler--DeWitt equation adds. For \(\kappa\gg1\) and small excitation the wave equation reduces, after removal of the vacuum phase, to a Caldirola--Kanai oscillator whose mass grows as \(\ee^{3u}\). A canonical dilation makes it time independent, and its Gaussian solutions are exact. %
In the ground state the guided configuration follows the shrinking packet, \(y_{\rm B}\propto\ee^{-3u/2}\), to leading order in the near-de Sitter approximation. At the packet centre the Hubble correction is \(h-1\simeq(\omega_\star'/2\nu)\ee^{-3u}\). This correction is the energy \(\tfrac12\omega_\star'\) of the zero-point motion divided by the cell action, and it is positive, because the wave function carries the energy of its own ground state into the Friedmann constraint. The quantum density \(Q\) that appears when this is decomposed into classical and quantum parts depends on where the guided configuration sits in the packet, \(Q\propto1-\xi^2/2\), while the sum does not, to \(9/(4\omega_\star'^2)\). We take this as the correct reading of the quantum potential in this model. It is a bookkeeping device that splits an energy, and the physical content is the total expansion rate. At the centre of the Gaussian packet, the leading adiabatic Bohmian quantum density equals the zero-point energy of the effective homogeneous oscillator per physical cell volume. For displaced Bohmian configurations, this density depends on the displacement parameter, and a formal Gaussian average gives half the centered value. Within the leading adiabatic Gaussian approximation, the centred quantum density scales as \(V^{-1}\propto a^{-3}\) and its effective pressure vanishes without oscillation averaging. The classical excitation instead behaves as dust after averaging rapid scalar oscillations. In the near-de Sitter, quadratic regime, its coherent-state occupation obeys Eq.~\eqref{eq:nbar}; for \(\kappa\gg1\) this gives \(\bar n\simeq[\nu\ee^{3u}/(2\kappa)]\overline{\rho}_{\rm exc}/\Lambda_\star\). The dark matter of the model is then a Bose condensate of the oscillation quanta of the field, with \(\bar n\sim10^{119}\) for a Hubble-sized cell, plus a zero-point term that is smaller by \(\bar n^{-1}\).

The numerical solutions of the full equation test these statements without the adiabatic truncation. The zero-point law is reproduced to \(10^{-3}\) for \(\xi=0\) and within \(10^{-2}\) for displaced configurations, with deviations that fall with the cell action as \(6\times10^{-3}\), \(3\times10^{-4}\) and \(10^{-4}\) at \(\nu=100\), \(300\) and \(1200\). For the excited condensate with \(\kappa=30\) the present Hubble rate approaches its classical value as \(290/\nu\) and \(296/\nu\) km s\(^{-1}\) Mpc\(^{-1}\), and the mean of \(Q\nu\ee^{3u}/\omega_\star'\) approaches unity, \(1.27\), \(1.12\) and \(1.05\) for \(\nu=300\), \(1200\) and \(4800\). For \(r=0.02\), where the oscillation is not separated from the expansion, the quantum corrections to \(h\), \(w_{\rm tot}\) and \(\Omega_Q\) all scale as \(1/\nu\) over \(100\le\nu\le450\). These results show that the quantum model has a controlled classical limit, in which the history is the classical one, and that the leading correction has the structure derived analytically. The physical value of \(\nu\) is of order \(10^{121}\), which makes \(\rho_Q/\Lambda_\star\sim\kappa/\nu\) unobservable, so the quantum density should be read as the finite-\(\nu\) completion of the classical dark sector and not as a prediction for present data. Its relevance is conceptual. It shows that in a Bohmian description the zero-point energy of the homogeneous mode of the field is cosmologically dust-like, and that this follows from the Wheeler--DeWitt dynamics and does not have to be imposed.

For the expansion history the content of the model is classical. The vacuum term gives \(w_{\rm DE}=-1\) exactly. In the near-de Sitter, quadratic approximation, the excitation has a relative energy modulation of amplitude \(3/(2\kappa)\) and period \(\pi/\omega_\star'\) in \(\ln(1+z)\). Over the full redshift interval, the evolving Hubble rate and anharmonicity also affect the phase and amplitude. The numerical comparisons reported here show deviations that decrease approximately as \(1/\kappa\), with coefficients \(3.0\), \(0.42\) and \(1.3\) for \(E(z)\), \(D_M\) and \(D_V\), respectively, for the specified normalization and phase sampling. The empirical \(D_V\) scaling gives an illustrative median one-percent similarity threshold near \(\kappa\gtrsim130\); it is not a fitted observational bound. The link between the vacuum oscillation scale and the critical structure of the potential remains encoded in \(\kappa^2\simeq3b/\delta\). The model cannot account for the time-dependent dark energy that DESI baryon acoustic oscillation data prefer at the level of a few standard deviations in combination with supernovae and the microwave background \cite{DESI2024,DESI}. If that preference persists, the vacuum sector would have to be generalized, for instance to a slowly rolling field, and the observation that the cuscuton coupling controls the hierarchy between vacuum and oscillation scales would remain the natural starting point of such an extension.

Several aspects of the construction are choices and not consequences, and we make them explicit because they define the scope of the claims. The ordering of the constraint is fixed, and the dependence of the Gaussian solution on it, including subleading changes to the amplitude and quantum density, is not explored. The guided configuration is not assigned a distribution, and the conserved current of a second-order equation does not provide the usual \(|\Psi|^2\) measure, so a typicality argument for the position of the configuration inside the packet is still needed, although the tested expansion rates were only weakly sensitive to it. The initial packet has the vacuum width, a displaced centre and a phase slope matched to the classical velocity, together with a derivative-corrected \(\Psi_u\) that is a definite rule and not a principle selecting the state. When the excitation energy is large compared with the vacuum energy the packet is strongly boosted, the quantum density is reduced by the factor \((F+Q)/h^2\), and the guided configuration can approach nodes of the wave function, where derivatives of the phase, and hence \(w_{\rm tot}\), are not controlled at the present resolution. In that regime the Hubble rate itself remains regular, and we therefore based the quantitative results on histories started at moderate excitation. Finally the model describes the homogeneous mode. Whether the condensate can cluster is decided by perturbations, whose quantum Jeans scale \(k_J\simeq a(16\pi G\rho)^{1/4}\mu_\star^{1/2}\) is a few times \(aH\) for \(\kappa=30\) and reaches galactic scales only for \(\kappa\sim10^{10}\) to \(10^{11}\), where the tuning of \(\delta\) is severe. The background results are independent of this requirement.

We have also established an approximate semiclassical correspondence between Bohmian Wheeler--DeWitt cosmology and the classical evolution of a homogeneous scalar field. It is shown that, in the large-action limit, the quantum corrections become suppressed and the Bohmian trajectories approach solutions of the classical Einstein–Klein–Gordon equations, as supported by both the WKB analysis and the numerical results. The section further demonstrates that coherent states of the effective harmonic oscillator obtained from the Wheeler–DeWitt equation provide a natural quantum description of classical scalar oscillations around the potential minimum. In the large-occupation-number regime, quantum fluctuations become small relative to the oscillation amplitude, allowing the coherent-state trajectory to acquire a semiclassical interpretation. Near the de Sitter minimum, averaging over the logarithmic scale factor is shown to be equivalent to averaging over cosmic time, provided the corresponding oscillation periods are consistently identified. In the rapid-oscillation regime, these averages yield a vanishing mean excitation pressure and an energy density that decreases inversely with the physical volume, reproducing the behavior of pressureless matter. The analysis therefore identifies the distinct roles of the large-action, large-occupation-number, and rapid-oscillation limits, and establishes how a classical dustlike scalar condensate can emerge as the coarse-grained semiclassical description of a coherent quantum state, while remaining distinct from the dustlike contribution associated with the Bohmian quantum potential.

In summary, one scalar field with a cuscuton-deformed potential supplies a cosmological constant, a condensate of dust-like oscillation quanta and a centred zero-point quantum density that is dust-like at leading adiabatic order. The parameter \(\delta\) controls the vacuum mass-to-Hubble hierarchy, while the quantum correction also depends on the cell action and the chosen state. The Bohmian Wheeler-DeWitt description turns the quantum state of the homogeneous mode into a definite contribution to the expansion rate, and the critical structure of the potential gives a compact setting in which the classical limit is known and the quantum corrections are solvable.

%%%%%%%%%%%%%%%%%%%%%%%%%%%%%%%%%%%%%%%%%%%%%%%%%%%%%%%%%%%%%%%%%%%%%%%%%%%%
\appendix
\section*{Appendices}

\section{Proof of the vacuum classification}
\label{app:vacuum}

Let \(0<b<3\), \(r\ge0\), \(C=\cosh x\), \(S=\sinh x\), \(k=\sqrt{b/3}\). For \(x<0\) every term of \(f_x\) in Eq.~\eqref{eq:f} is nonpositive and the first is strictly negative, so there is no stationary point. At \(r=0\) the unique minimum is \(x=0\). For \(x>0\), \(f_x=3\mathcal F(x,r)\) with
\begin{equation}
    \mathcal F=xr^2-(C+xS)r+(1-k^2)SC,\qquad r_\pm(x)=\frac{C+xS\pm\sqrt\Delta}{2x},\quad\Delta=(C-xS)^2+4k^2xSC>0 .
\end{equation}
Since \(\mathcal F(x,0)>0\) and \(\mathcal F(x,S)=\mathcal F(x,C/x)=-k^2SC<0\), one has \(0<r_-<\min(S,C/x)\) and \(r_+>\max(S,C/x)\). At a stationary point the elimination of \(1-k^2\) gives \(\mathcal F_x=r\,L(r)\) with \(L=1/S+xS^2/C+r[1-x(C^2+S^2)/(SC)]\), whose coefficient of \(r\) is negative because \(x(C^2+S^2)/(SC)=2x\coth2x>1\). If \(xS\le C\) then \(L(S)=C(C/S-x)\ge0\), and if \(xS\ge C\) then \(L(C/x)=(xS-C)^2/(xC)\ge0\). With \(r_-<\min(S,C/x)\) this gives \(\mathcal F_x(x,r_-)>0\), hence \(r_-'(x)=\mathcal F_x/\sqrt\Delta>0\). The lower root increases from \(0\) to infinity, so every \(r>0\) has exactly one stationary point on this branch, and there the curvature is positive.

The potential factorizes as \(f=\tfrac32h_+h_-\) with \(h_\pm=C-rx\pm kS\). For \(x>0\), \(h_-=x[g(x)-r]\) with \(g=(C-kS)/x\), which diverges at both ends of the half-line. One has \(g'=-G/x^2\) with \(G=C(1+kx)-S(x+k)\), \(G(0)=1\), \(G'=x(kS-C)<0\) and \(G\to-\infty\), so \(g\) has a unique minimum \(\rc=g(x_c)\). For \(r<\rc\) both \(h_\pm\) are positive for \(x>0\), and for \(x<0\) positivity follows from \((C-rx)^2-k^2S^2\ge1+(1-k^2)S^2>0\). Let \(r_0=\min_{x>0}C/x\). Since \(k>0\), \(\rc<r_0\le C/x<r_+\), so there are no upper-branch stationary points for \(0\le r\le\rc\) and the lower-branch point is the unique global minimum. At \(r=\rc\) one has \(h_-=h_-'=0\) at \(x_c\), \(f(x_c)=0\) and \(f''(x_c)=3k\sinh x_c(\cosh x_c-k\sinh x_c)>0\). Along the lower branch the envelope relation gives \(\dd f_\star/\dd r=f_r=-3x_\star(C-rx_\star)<0\), so \(f_\star\) is positive below \(\rc\) and negative above it. Finally \(\rc x_c=C-kS\) at criticality gives \(-\rc f_\star'(\rc)=f''(x_c)\), and with \(\delta=1-r/\rc\) one finds \(f_\star=f''(x_c)\delta+O(\delta^2)\), \(f''_\star=f''(x_c)+O(\delta)\) and therefore \(\kappa^2=3b/\delta+O(1)\). At the boundary itself \(H_\star=0\) and the normalization of the quantum problem fails.

\section{Exact Gaussian solutions of the transformed oscillator}
\label{app:gaussian}

Let \(\alpha=1/m_0\), \(\gamma=m_0\kappa^2\) and \(\beta=3/2\), so that \(K=\tfrac12[\alpha P^2+\gamma q^2+\beta(qP+Pq)]\) with \(P=-\ii\partial_q\). Acting on \(\psi=\exp[-\tfrac12\Gamma q^2]\) one finds \(P\psi=\ii\Gamma q\psi\), \(P^2\psi=(\Gamma-\Gamma^2q^2)\psi\) and \((qP+Pq)\psi=(2\ii\Gamma q^2-\ii)\psi\), hence
\begin{equation}
    2K\psi=\left[\alpha\Gamma-\ii\beta+(\gamma-\alpha\Gamma^2+2\ii\beta\Gamma)q^2\right]\psi .
\end{equation}
For a stationary state the coefficient of \(q^2\) must vanish, \(\alpha\Gamma^2-2\ii\beta\Gamma-\gamma=0\), and the root with positive real part is \(\Gamma=w=(\ii\beta+\sqrt{\alpha\gamma-\beta^2})/\alpha=m_0(\omega_\star'+\tfrac32\ii)\), using \(\alpha\gamma=\kappa^2\) and \(\kappa^2-\beta^2=\omega_\star'^2\). The eigenvalue is \(2E_0=\alpha w-\ii\beta=\omega_\star'\). The time-dependent problem \(\ii\psi_u=K\psi\) with \(\psi=\exp[-\tfrac12\Gamma(u)q^2+c(u)]\) gives \(\Gamma'=-\ii\alpha\Gamma^2+\ii\gamma-2\beta\Gamma\). Linearizing around the fixed point, \(\delta\Gamma'=-2\ii\omega_\star'\,\delta\Gamma\), so deviations oscillate with the frequency \(2\omega_\star'\) and do not decay. This is the breathing mode seen in the numerical solutions.

For a coherent state \(\psi=\psi_0(q-q_c)\exp[\ii P_c(q-q_c)+\ii\gamma_c]\) the classical orbit satisfies \(\dot q_c=\alpha P_c+\beta q_c\) and \(\dot P_c=-\gamma q_c-\beta P_c\), with the dot denoting \(\dd/\dd u\). The phase is \(\gamma_c=\int(P_c\dot q_c-K_{\rm cl})\dd u-E_0u\) with \(K_{\rm cl}=K(q_c,P_c)\). The Bohmian velocity of a particle in the \(q\) frame is \(\dot q=\alpha\,\Imag(\partial_q\ln\psi)+\beta q=\alpha[P_c-\Imag(w)(q-q_c)]+\beta q\), and since \(\alpha\Imag w=\beta\) this equals \(\alpha P_c+\beta q_c=\dot q_c\). The offset \(q-q_c\) is therefore constant, which is the statement used in Sec.~\ref{sec:adiabatic}.

\section{Numerical scheme and initial data}
\label{app:numerics}

On the interval \([x_L,x_R]\) of length \(L_x\) the field is discretized on the grid \(x_j=x_L+jL_x/(N+1)\), \(j=1,\dots,N\), with the orthonormal discrete sine transform, so that \(\partial_x^2\to-k_j^2\) with \(k_j=\pi j/L_x\), and the potential term is a diagonal multiplication on the grid. The state is the vector of sine coefficients of \(\chi\) and \(\chi_s\), and \(\chi\), \(\chi_x\), \(\chi_s\), \(\chi_{sx}\), \(\chi_{ss}\) at the Bohmian position are evaluated from the series. The integration uses DOP853 with the stated tolerances and is restarted every \(100\nu/300\) units of \(s\), and the diagnostics are computed at a regular output spacing of \(0.05\,\nu/300\) in \(s\) for the excited runs. Spectral resolution is controlled by \(k_{\max}=N\pi/L_x\), which has to exceed the largest guided momentum \(\nu\ee^{3u}|v|/c_x^2\) with a margin of about a factor of two, and the transition width of the cutoff has to be resolved by the grid.

The initial packet is \(\Psi_{\rm i}=\mathcal N\exp[-(x-x_{\rm i})^2/(4\sigma_{\rm i}^2)+\ii p_{\rm i}(x-x_{\rm i})]\,W(x)\) with \(p_{\rm i}=\nu v_{\rm i}/c_x^2\). The cutoff \(W\) equals \(1\) in the interior of \([l_c,r_c]\) and vanishes at its ends, with transition width \(m=0.1\) given by \(W=\ee^{-1/(1-t)}/[\ee^{-1/t}+\ee^{-1/(1-t)}]\) for \(0<t<1\), where \(t=\max[(l_c+m-x)/m,(x-r_c+m)/m]\). The width \(\sigma_{\rm i}\) is the ground-state width of the auxiliary oscillator \(\hat H_\phi=-\partial_{\delta\phi}^2/(2V_{\rm i})+\tfrac12V_{\rm i}\mu_\star^2(\delta\phi)^2\), with \(\delta\phi=\phi-\phi_\star\) expressed in \(x\), which is a convenient localized family and is not a derived vacuum of the coupled gravitational system. The packet is centred on the classical reference position and the guided configuration is placed at its centre. The normalization of \(\Psi\) does not enter the guidance equations.

A packet alone does not define a solution of the second-order equation, and \(\Psi_u\) must also be supplied. Let \(K(u)=-c_x^2\partial_x^2+\nu^2\ee^{6u}F\) and \(\Omega=K^{1/2}\), both positive on the grid. A first-order operator \(A\) with \(\Psi_u=A\Psi\) would factorize the equation exactly if \(A'+A^2+K=0\). The choice \(A=D-\ii\Omega\) with \(\{\Omega,D\}=-\Omega'\) removes the imaginary part of the residual and leaves \(D'+D^2\). We use the corrected operator
\begin{equation}
    A_3=D+D_2-\ii(\Omega+C),
\end{equation}
where the real symmetric matrices satisfy \(\{\Omega,C\}=D'+D^2\) and \(\{\Omega,D_2\}=-(C'+\{D,C\})\), with \(\{A,B\}=AB+BA\) and primes denoting \(u\) derivatives. This cancels successive terms of the residual and leaves \(D_2'+\{D,D_2\}+D_2^2-C^2-\ii\{D_2,C\}\). The derivatives of \(\Omega\) follow from \(\Omega^2=K\) through \(\{\Omega,\Omega'\}=K'\), \(\{\Omega,\Omega''\}=K''-2\Omega'^2\) and \(\{\Omega,\Omega'''\}=K'''-3\{\Omega',\Omega''\}\) with \(K'=6\nu^2\ee^{6u}F\), \(K''=6K'\) and \(K'''=36K'\), and those of \(D\) and \(C\) from the differentiated equations, \(\{\Omega,D'\}=-\Omega''-\{\Omega',D\}\), \(\{\Omega,D''\}=-\Omega'''-\{\Omega'',D\}-2\{\Omega',D'\}\) and \(\{\Omega,C'\}=R_1'-\{\Omega',C\}\) with \(R_1=D'+D^2\). In the eigenbasis of \(K\) at \(u=0\), with eigenvalues \(\omega_j^2\), each equation \(\{\Omega,X\}=Y\) is solved by \(X_{jk}=Y_{jk}/(\omega_j+\omega_k)\), and the positive denominator makes the finite-dimensional solution unique. After this single use the second-order equation is integrated without further approximation. The preparation is a definite rule that is not an exact positive-frequency factorization and does not select a unique state. For the ground-state tests of Sec.~\ref{sec:numerics} the initial \(\Psi_u\) is instead the analytic derivative of Eq.~\eqref{eq:Psi0}, \(\Psi_u=[-\ii\nu-\ii E_0-\tfrac34-\tfrac32wy^2]\Psi\) at \(u=0\).

\section{Small-coupling expansion of the minimum}
\label{app:small}

For small \(r\) write the potential as \(f=\tfrac12[(3-b)\cosh^2x+b]-3rx\cosh x+\tfrac32r^2x^2\). The stationary condition \(f_x=(3-b)\sinh x\cosh x-3r(\cosh x+x\sinh x)+3r^2x=0\) gives at leading order \((3-b)x-3r=0\), and since the equation is odd under \((x,r)\to(-x,-r)\) the solution is an odd function of \(r\), \(x_\star=3r/(3-b)+O(r^3)\). Keeping terms through \(x^2\) in the first part and through \(x\) in the second, \(f(x)\simeq\tfrac32+\tfrac12(3-b)x^2-3rx\), and at \(x_\star\)
\begin{equation}
    f_\star=\tfrac32+\frac{9r^2}{2(3-b)}-\frac{9r^2}{3-b}=\tfrac32-\frac{9r^2}{2(3-b)}+O(r^4).
\end{equation}
The curvature is \(f_{xx}=(3-b)(\cosh^2x+\sinh^2x)-3r(2\sinh x+x\cosh x)+3r^2\). With \(\cosh^2x+\sinh^2x=1+2x^2+O(x^4)\) and \(2\sinh x+x\cosh x=3x+O(x^3)\),
\begin{equation}
    f_{xx}(x_\star)=(3-b)+\frac{18r^2}{3-b}-\frac{27r^2}{3-b}+3r^2=(3-b)-\frac{3b\,r^2}{3-b}+O(r^4),
\end{equation}
and the third derivative \(f_{xxx}=4(3-b)\sinh x\cosh x-3r(3\cosh x+x\sinh x)\) equals \(12r-9r=3r\) at the minimum. In terms of \(\phi\) these results reproduce \(\phi_\star=3\alpha_{\rm c}/[AB^2(3-B^2)]\), \(V_\star=\tfrac32A^2-9\alpha_{\rm c}^2/[2B^2(3-B^2)]\) and \(V_1''(\phi_\star)=A^2B^2(3-B^2)-3B^2\alpha_{\rm c}^2/(3-B^2)\), and for \(b=3/2\) and \(r=0.02\) they give \(f_\star=1.4988\) and \(f_{xx}=1.4988\) against the exact values \(1.49880\) and \(1.49880\). The expansion parameter is \(|x_\star|=3r/(3-b)\), so it fails well before \(r\) approaches \(\rc\), where \(x_\star\to x_c=1.609\).

\section{Equation of motion and Raychaudhuri equation along Bohmian trajectories}
\label{app:bohm}

Write the effective potential as \(\mathcal F=F+Q\), so that Eq.~\eqref{eq:HJ} reads \(S_u^2-c_x^2S_x^2=\nu^2\ee^{6u}\mathcal F\). Differentiating with respect to \(x\) gives \(S_uS_{ux}-c_x^2S_xS_{xx}=\tfrac12\nu^2\ee^{6u}\mathcal F_x\). Along the trajectory \(\dd x/\dd u=v/h=-c_x^2S_x/S_u\) by Eq.~\eqref{eq:guidance}, so
\begin{equation}
    \frac{\dd S_x}{\dd u}=S_{xu}+S_{xx}\frac{\dd x}{\dd u}=\frac{S_uS_{ux}-c_x^2S_xS_{xx}}{S_u}=\frac{\nu^2\ee^{6u}\mathcal F_x}{2S_u}=-\frac{\nu\ee^{3u}\mathcal F_x}{2h},
\end{equation}
where \(S_u=-\nu\ee^{3u}h\) has been used. Since \(v=c_x^2\ee^{-3u}S_x/\nu\) one obtains
\begin{equation}
    \frac{\dd v}{\dd u}=-3v-\frac{c_x^2\mathcal F_x}{2h}=-3v-\frac{3b\,(F_x+Q_x)}{h},
\end{equation}
which is Eq.~\eqref{eq:bohm-eom}. For the Hubble rate, differentiate \(h^2=v^2/c_x^2+F+Q\) along the trajectory. The three contributions are \(\dd(v^2/c_x^2)/\dd u=(2v/c_x^2)\,\dd v/\dd u=-6K-(v/h)(F_x+Q_x)\) since \(2\cdot3b/c_x^2=1\), \(\dd F/\dd u=F_xv/h\), and \(\dd Q/\dd u=(\partial_uQ)_x+Q_xv/h\). The terms in \(F_x\) and \(Q_x\) cancel in the sum, leaving \(2h\,\dd h/\dd u=-6K+(\partial_uQ)_x\), which is Eq.~\eqref{eq:raych}. Substituting into \(w_{\rm tot}=-1-2h^{-1}(\dd h/\dd u)/3\) gives \(w_{\rm tot}=(K-F-Q-\tfrac13(\partial_uQ)_x)/h^2=(K-F+\Pi_Q)/h^2\) with \(\Pi_Q\) of Eq.~\eqref{eq:PiQ}, after the constraint is used to eliminate \(h^2\).

\section*{Acknowledgements}

AB thanks the Indian Institute of Technology Kanpur for institutional support and research facilities. RS thanks the Council of Scientific and Industrial Research, Government of India, for financial support through the CSIR RA scheme.


\begin{thebibliography}{99}
\bibitem{Weinberg1989} S.~Weinberg, The cosmological constant problem, Rev.\ Mod.\ Phys.\ \textbf{61}, 1 (1989).
\bibitem{Martin2012} J.~Martin, Everything you always wanted to know about the cosmological constant problem (but were afraid to ask), C.\ R.\ Physique \textbf{13}, 566 (2012).
\bibitem{Planck2018} N.~Aghanim \textit{et al.} (Planck Collaboration), Planck 2018 results. VI. Cosmological parameters, Astron.\ Astrophys.\ \textbf{641}, A6 (2020).
\bibitem{DESI2024} A.~G.~Adame \textit{et al.} (DESI Collaboration), DESI 2024 VI: cosmological constraints from the measurements of baryon acoustic oscillations, J.\ Cosmol.\ Astropart.\ Phys.\ \textbf{02}, 021 (2025).
\bibitem{DESI} M.~Abdul-Karim \textit{et al.} (DESI Collaboration), DESI DR2 results II: measurements of baryon acoustic oscillations and cosmological constraints, Phys.\ Rev.\ D \textbf{112}, 083515 (2025).
\bibitem{Kamenshchik2001} A.~Kamenshchik, U.~Moschella and V.~Pasquier, An alternative to quintessence, Phys.\ Lett.\ B \textbf{511}, 265 (2001).
\bibitem{Bilic2002} N.~Bili\'c, G.~B.~Tupper and R.~D.~Viollier, Unification of dark matter and dark energy: the inhomogeneous Chaplygin gas, Phys.\ Lett.\ B \textbf{535}, 17 (2002).
\bibitem{Bento2002} M.~C.~Bento, O.~Bertolami and A.~A.~Sen, Generalized Chaplygin gas, accelerated expansion and dark-energy--matter unification, Phys.\ Rev.\ D \textbf{66}, 043507 (2002).
\bibitem{Scherrer2004} R.~J.~Scherrer, Purely kinetic k-essence as unified dark matter, Phys.\ Rev.\ Lett.\ \textbf{93}, 011301 (2004).
\bibitem{Bertacca2010} D.~Bertacca, N.~Bartolo and S.~Matarrese, Unified dark matter scalar field models, Adv.\ Astron.\ \textbf{2010}, 904379 (2010).
\bibitem{Sahni2000} V.~Sahni and L.~Wang, New cosmological model of quintessence and dark matter, Phys.\ Rev.\ D \textbf{62}, 103517 (2000).
\bibitem{Matos2000} T.~Matos and L.~A.~Ure\~na-L\'opez, Quintessence and scalar dark matter in the universe, Class.\ Quantum Grav.\ \textbf{17}, L75 (2000).
\bibitem{Arbey2006} A.~Arbey, Dark fluid: a complex scalar field to unify dark energy and dark matter, Phys.\ Rev.\ D \textbf{74}, 043516 (2006).
\bibitem{Turner1983} M.~S.~Turner, Coherent scalar-field oscillations in an expanding universe, Phys.\ Rev.\ D \textbf{28}, 1243 (1983).
\bibitem{Hu2000} W.~Hu, R.~Barkana and A.~Gruzinov, Fuzzy cold dark matter: the wave properties of ultralight particles, Phys.\ Rev.\ Lett.\ \textbf{85}, 1158 (2000).
\bibitem{Hui2017} L.~Hui, J.~P.~Ostriker, S.~Tremaine and E.~Witten, Ultralight scalars as cosmological dark matter, Phys.\ Rev.\ D \textbf{95}, 043541 (2017).
\bibitem{Marsh2016} 
D.~J.~E.~Marsh, Axion cosmology, Phys.\ Rep.\ \textbf{643}, 1 (2016).
\bibitem{Ferreira2021} E.~G.~M.~Ferreira, Ultra-light dark matter, Astron.\ Astrophys.\ Rev.\ \textbf{29}, 7 (2021).
\bibitem{Johnson2008} M.~C.~Johnson and M.~Kamionkowski, Dynamical and gravitational instability of oscillating-field dark energy and dark matter, Phys.\ Rev.\ D \textbf{78}, 063010 (2008).
\bibitem{Hlozek2015} R.~Hlo\v{z}ek, D.~Grin, D.~J.~E.~Marsh and P.~G.~Ferreira, A search for ultralight axions using precision cosmological data, Phys.\ Rev.\ D \textbf{91}, 103512 (2015).
\bibitem{Irsic2017} V.~Ir\v{s}i\v{c}, M.~Viel, M.~G.~Haehnelt, J.~S.~Bolton and G.~D.~Becker, First constraints on fuzzy dark matter from Lyman-\(\alpha\) forest data and hydrodynamical simulations, Phys.\ Rev.\ Lett.\ \textbf{119}, 031302 (2017).
\bibitem{Rogers2021} K.~K.~Rogers and H.~V.~Peiris, Strong bound on canonical ultralight axion dark matter from the Lyman-alpha forest, Phys.\ Rev.\ Lett.\ \textbf{126}, 071302 (2021).
\bibitem{DeWitt1967} B.~S.~DeWitt, Quantum theory of gravity. I. The canonical theory, Phys.\ Rev.\ \textbf{160}, 1113 (1967).
\bibitem{Hartle1983} J.~B.~Hartle and S.~W.~Hawking, Wave function of the universe, Phys.\ Rev.\ D \textbf{28}, 2960 (1983).
\bibitem{Vilenkin1984} A.~Vilenkin, Quantum creation of universes, Phys.\ Rev.\ D \textbf{30}, 509 (1984).
\bibitem{Kiefer1988} C.~Kiefer, Wave packets in minisuperspace, Phys.\ Rev.\ D \textbf{38}, 1761 (1988).
\bibitem{Halliwell1991} J.~J.~Halliwell, Introductory lectures on quantum cosmology, in \textit{Quantum Cosmology and Baby Universes}, World Scientific (1991), arXiv:0909.2566.
\bibitem{Kiefer2012} C.~Kiefer, \textit{Quantum Gravity}, 3rd ed., Oxford University Press (2012).
\bibitem{Hawking1986} S.~W.~Hawking and D.~N.~Page, Operator ordering and the flatness of the universe, Nucl.\ Phys.\ B \textbf{264}, 185 (1986).
\bibitem{Bohm1952} D.~Bohm, A suggested interpretation of the quantum theory in terms of ``hidden'' variables. I, Phys.\ Rev.\ \textbf{85}, 166 (1952).
\bibitem{Holland1993} P.~R.~Holland, \textit{The Quantum Theory of Motion}, Cambridge University Press (1993).
\bibitem{Durr1992} D.~D\"urr, S.~Goldstein and N.~Zangh\`i, Quantum equilibrium and the origin of absolute uncertainty, J.\ Stat.\ Phys.\ \textbf{67}, 843 (1992).
\bibitem{Colistete2000} R.~Colistete Jr., J.~C.~Fabris and N.~Pinto-Neto, Gaussian superpositions in scalar-field quantum cosmology, Phys.\ Rev.\ D \textbf{62}, 083507 (2000).
\bibitem{Peter2007} P.~Peter, E.~J.~C.~Pinho and N.~Pinto-Neto, A non-inflationary model with scale invariant cosmological perturbations, Phys.\ Rev.\ D \textbf{75}, 023516 (2007).
\bibitem{Peter2008} P.~Peter and N.~Pinto-Neto, Cosmology without inflation, Phys.\ Rev.\ D \textbf{78}, 063506 (2008).
\bibitem{PintoNeto2013} N.~Pinto-Neto and J.~C.~Fabris, Quantum cosmology from the de Broglie--Bohm perspective, Class.\ Quantum Grav.\ \textbf{30}, 143001 (2013).
\bibitem{Kiefer2022} C.~Kiefer and P.~Peter, Time in quantum cosmology, Universe \textbf{8}, 36 (2022).
\bibitem{Falciano2007} F.~T.~Falciano, N.~Pinto-Neto and E.~S.~Santini, Inflationary nonsingular quantum cosmological model, Phys.\ Rev.\ D \textbf{76}, 083521 (2007).
\bibitem{Demaerel2019} T.~Demaerel and W.~Struyve, Elimination of cosmological singularities in quantum cosmology by suitable operator orderings, Phys.\ Rev.\ D \textbf{100} (2019), arXiv:1904.09244.
\bibitem{Bqc1} S.~Basilakos, G.~Kouniatalis, E.~N.~Saridakis and C.~Tzerefos, Bohmian quantum cosmology from the Wheeler--DeWitt equation, Phys.\ Lett.\ B \textbf{875}, 140339 (2026).
\bibitem{Afshordi2006} N.~Afshordi, D.~J.~H.~Chung, M.~Kleban and G.~Geshnizjani, Cuscuton cosmology: dark energy meets modified gravity, Phys.\ Rev.\ D \textbf{75}, 021302(R) (2007).
\bibitem{Afshordi2007} N.~Afshordi, D.~J.~H.~Chung and G.~Geshnizjani, Cuscuton: a causal field theory with an infinite speed of sound, Phys.\ Rev.\ D \textbf{75}, 083513 (2007).
\bibitem{CuscDE} D.~Bazeia, J.~D.~Dantas and S.~Santos da Costa, Cuscuton-like contribution to dark energy evolution, Eur.\ Phys.\ J.\ C \textbf{85}, 196 (2025).
\bibitem{Kanai1948} E.~Kanai, On the quantization of the dissipative systems, Prog.\ Theor.\ Phys.\ \textbf{3}, 440 (1948).
\bibitem{Dekker1981} H.~Dekker, Classical and quantum mechanics of the damped harmonic oscillator, Phys.\ Rep.\ \textbf{80}, 1 (1981).
\bibitem{Lewis1969} H.~R.~Lewis Jr.\ and W.~B.~Riesenfeld, An exact quantum theory of the time-dependent harmonic oscillator and of a charged particle in a time-dependent electromagnetic field, J.\ Math.\ Phys.\ \textbf{10}, 1458 (1969).
\bibitem{Hairer1993} E.~Hairer, S.~P.~N\o rsett and G.~Wanner, \textit{Solving Ordinary Differential Equations I: Nonstiff Problems}, Springer (1993).
\bibitem{Boyd2001} J.~P.~Boyd, \textit{Chebyshev and Fourier Spectral Methods}, 2nd ed., Dover (2001).
\bibitem{Eisenstein2005} D.~J.~Eisenstein \textit{et al.} (SDSS Collaboration), Detection of the baryon acoustic peak in the large-scale correlation function of SDSS luminous red galaxies, Astrophys.\ J.\ \textbf{633}, 560 (2005).

\end{thebibliography}
\end{document}